\documentclass[letterpaper,twocolumn,10pt]{article}
\usepackage{usenix2024_SOUPS}

\usepackage{tikz}
\usepackage{amsmath}
\usepackage{amssymb}
\usepackage{booktabs}
\usepackage{cite}
\usepackage{colortbl}
\usepackage{graphicx}
\usepackage{float}
\usepackage{tabularx}
\usepackage{textcomp}
\usepackage{xcolor}
\usepackage{caption}
\usepackage{subcaption}
\usepackage{listings}
\usepackage{comment}
\usepackage{url}

\usepackage{enumitem}
\usepackage{nopageno}

\usepackage{titlesec}
\titlespacing*{\section}{0pt}{10pt}{6pt}
\titlespacing*{\subsection}{0pt}{8pt}{4pt}
\titlespacing*{\subsubsection}{0pt}{4pt}{2pt}

\usepackage{filecontents}
\renewcommand{\paragraph}[1]{\vspace*{2pt}\noindent\textbf{#1}}
\newcommand{\remove}[1]{}

\begin{document}

\date{}

\title{\Large \bf Evaluating the Usability of Differential Privacy Tools with Data Practitioners}

\def\plainauthor{Author name(s) for PDF metadata. Don't forget to anonymize for submission!}

\author{
{\rm Ivoline C. Ngong}\\
University of Vermont
\and
{\rm Brad Stenger}\\
University of Vermont
\and
{\rm Joseph P. Near}\\
University of Vermont
\and
{\rm Yuanyuan Feng}\\
University of Vermont
} 

\maketitle

\begin{abstract}
Differential privacy (DP) has become the gold standard in privacy-preserving data analytics, but implementing it in real-world datasets and systems remains challenging. Recently developed DP tools aim to make DP implementation easier, but limited research has investigated these DP tools' usability. Through a usability study with 24 US data practitioners with varying prior DP knowledge, we evaluated the usability of four open-source Python-based DP tools: DiffPrivLib, Tumult Analytics, PipelineDP, and OpenDP.
Our study results suggest that these DP tools moderately support data practitioners' DP understanding and implementation; that Application Programming Interface (API) design and documentation are vital for successful DP implementation and user satisfaction. We provide evidence-based recommendations to improve DP tools' usability to broaden DP adoption.
\end{abstract}

\section{Introduction}

Advances in big data analytics have propelled the collection and processing of massive amounts of data, including sensitive data such as medical records, financial information, and other personally identifiable information.
The analysis of this sensitive data may result in the accidental leakage of individuals' data~\cite{databreach1, databreach2}, even when anonymization techniques are used~\cite{shokri2017membership, jayaraman2020revisiting, carlini2021extracting, carlini2019secret}.
Differential privacy (DP) can mitigate these risks~\cite{dwork2006calibrating, dwork2014algorithmic} by guaranteeing the results of statistical analyses will not reveal too much personal information about any individual. 
By adding carefully calibrated noise to data, DP protects sensitive data while still revealing high-level statistical insights.
Due to its tremendous potential to revolutionize privacy-preserving data analysis, DP attracts considerable research~\cite{dwork2014algorithmic}. Leading government organizations and technology companies, including the U.S. Census Bureau~\cite{censusdpuse},  Google~\cite{googledpuse}, Apple~\cite{appledpuse}, and Microsoft~\cite{microsoftdpuse} have also adopted DP to protect individuals' data privacy.

However, current DP adoption is limited outside of large organizations and companies~\cite{desfontaines2020lowering}, primarily because implementing DP from scratch is complex and error-prone~\cite{kifer2020guidelines}.
DP implementations must carefully account for the privacy budget, generate appropriate random noise, and require systems to be safe against known side-channel vulnerabilities. Additionally, scaling these systems to real-world datasets often requires significant software engineering effort.

To address these challenges, various tools, frameworks, and libraries~\cite{opendp, dpcreator, diffprivlib, googledp, pipelinedp, gaboardi2016psi, privacybeam, zetasql, wilson2020differentially, chorusweb, johnson2020chorus, mcsherry2009privacy, yousefpour2021opacus, papernot2019machine} (collectively called ``DP tools'' hereafter) have been developed to make DP implementation accessible to \textbf{data practitioners} --- defined in this paper as professionals who have data analysis and programming skills but may not be familiar with DP. 
These DP tools intend to help data practitioners implement DP solutions without privacy failures.
Currently, no research has systematically evaluated the usability of these DP tools; therefore, it remains unclear if they truly enable data practitioners to effectively implement DP solutions. If not, usability may be the bottleneck for wider DP adoption.

In this study, we have assessed four open-source Python-based DP tools through a mixed-methods usability study with 24 US data practitioners, evaluating four widely-used usability criteria--learnability, efficiency, error prevention, and user satisfaction~\cite{nielsen1996usability} -- to investigate three research questions:
\begin{itemize}[itemsep=-1pt,topsep=1pt]
\item How effectively can DP tools help data practitioners understand DP concepts? (RQ1: DP Understanding)
\item How effectively can DP tools help data practitioners implement DP solutions? (RQ2: DP Implementation) 
\item How satisfied are data practitioners with DP tools for their DP implementation? (RQ3: User Satisfaction)
\end{itemize}
We conducted the first comprehensive cross-tool usability study of four Python-based DP tools with data practitioners. The focus on data practitioners---the potential adopters of DP---enriches the currently end user-centered DP user research. Our contribution lies in the identification of these DP tools' usability issues and in our recommendations to improve DP tools' usability to facilitate broader DP adoption.
\section{Related Work}

\paragraph{DP and Implementation Challenges.}
Differential privacy (DP)~\cite{dwork2006calibrating, dwork2014algorithmic}  is a formal privacy definition designed to allow statistical analysis while protecting information about individuals. Differentially private analyses, often called \emph{mechanisms}, typically add random noise to analysis results in order to achieve privacy.
Formally, two datasets $D, D' \in \mathcal{D}$ are called \emph{neighboring datasets} if they differ in one person's data, and a mechanism $\mathcal{M}$ satisfies $(\epsilon, \delta)$-DP if for all neighboring datasets $D$ and $D'$ and sets of outcomes $S$:
\vspace*{-7pt}
\[ \Pr[\mathcal{M}(D) \in S] \leq e^\epsilon \Pr[\mathcal{M}(D') \in S] + \delta \]

\vspace*{-7pt} \noindent The $\epsilon$ parameter is the \emph{privacy parameter} or \emph{privacy budget}; a smaller $\epsilon$ results in stronger privacy, while a larger $\epsilon$ results in weaker privacy.
Noise drawn from the Laplace or Gaussian distributions can be used to achieve differential privacy.



\paragraph{Existing DP Tools.}
Implementing DP mechanisms is challenging. Data practitioners must determine the amount of noise to add, limit the total privacy budget, and ensure the system is free of common DP bugs~\cite{mironov2012significance, haney2022precision, casacuberta2022widespread, jin2022we}). 
Numerous tools and libraries have attempted to make implementing DP easier for data practitioners~\cite{opendp, dpcreator, diffprivlib, googledp, pipelinedp, gaboardi2016psi, privacybeam, zetasql, wilson2020differentially, chorusweb, johnson2020chorus, mcsherry2009privacy, yousefpour2021opacus, papernot2019machine}, often by handling the tricky parts of DP automatically. For example, tools may calculate sensitivity automatically~\cite{zetasql, chorusweb, mcsherry2009privacy} and ensure the privacy budget is not violated~\cite{opendp, dpcreator, pipelinedp, gaboardi2016psi, wilson2020differentially, johnson2020chorus, mcsherry2009privacy, yousefpour2021opacus, papernot2019machine}. They may provide vetted implementations of basic DP mechanisms like the Laplace mechanism~\cite{opendp, diffprivlib, googledp, yousefpour2021opacus, papernot2019machine}, and some also support machine learning applications~\cite{yousefpour2021opacus, papernot2019machine, diffprivlib}.
Notably, DPCreator~\cite{dpcreator} and Private data Sharing Interface (PSI)~\cite{gaboardi2016psi} provide graphical interfaces designed for non-experts; the other tools require data science knowledge but reduce the need for DP expertise.


\paragraph{User Research around DP Understanding.}
Existing user research around DP understanding mostly focuses on end users, whose data would be in a differentially private dataset.
Bullek et al. \cite{bullek2017towards} examined if animated spinners can effectively communicate DP privacy guarantees to end users. Their participants preferred spinners with higher privacy levels but did not fully trust the spinners.
Cummings et al. \cite{cummings2021need} studied how various DP explanations impact end-user perceptions. They found DP explanations raised participants' expectations of privacy, but did not increase their willingness to share data.
Other studies explored how to better communicate DP concepts to end users.
Xiong et al. \cite{xiong2020towards} assessed the use of scenarios to communicate the privacy guarantees of three different DP models with participants from the USA and India.
Kühtreiber et al.\cite{kuhtreiber2022replication} replicated this study with participants from Germany. Both studies indicated that end users lack understanding of DP and highlighted a need for more effective DP communication. German participants were more willing to share data compared to those in the USA and India.
Ashena et al.\cite{ashenaSP2024} also found interactive visual tools helped communicate the trade-off between accuracy and privacy loss.
These studies suggest that end users have difficulty understanding DP and reservations towards DP's privacy protection.

Currently, limited user research has examined the perspective of data practitioners, who particularly need adequate DP understanding to correctly implement it.
One notable study by Nanayakkara et al.\cite{nanayakkara2022visualizing} tested an interactive interface called Visualizing Privacy (ViP) with data practitioners without DP background, and found visualizing relationships between $\epsilon$, accuracy, and disclosure risk helped them judge DP noise. 
Our study extends this line of research to investigate if DP tools could assist data practitioners' DP understanding.

\paragraph{Usability around DP Implementation.}
Garrido et al. \cite{garrido2022lessons} interviewed 24 privacy practitioners and identified both organizational and technical challenges for DP implementation in the industry.
Their findings suggested that API-based DP tools could streamline data access integration and DP implementation across the enterprise.

A few studies have evaluated the usability of specific DP tools.
Murtagh et al.~\cite{murtagh2018usable} studied the usability of the web-based DP tool Privacy-preserving Integration (PSI) tool. Study participants succeeded at assigned tasks, but also identified areas of confusion and error.
Sarathy et al.~\cite{sarathy2023don} conducted a usability study with 19 non-expert participants using the DP Creator prototype to understand perceptions, challenges, and opportunities around DP analysis. Their findings highlight user challenges including users' poor understanding of decision implications, and difficulty accessing raw data and managing workflows. 
%
We expand prior research to evaluate the usability of multiple DP tools with data practitioners.

Recently, Govtech Singapore conducted a usability assessment of DP tools~\cite{govtechbenchmarks}. They compared the same four Python-based DP tools' capabilities in analysis, security, usability, and differential privacy, generating a usability benchmark for these tools. This was an expert heuristic review without user testing, which can be subjective and lacks depth compared to our usability study that involves data practitioners. 

\paragraph{Usability of Non-DP Tools.} 
While our study focuses on DP tools' usability, it is critical to draw implications from prior research on non-DP tools.
There is usability research on non-DP data science tools that require programming skills.
Akil et al.~\cite{akil2017usability} compared the usability of three prominent distributed data processing platforms for cloud computing (MapReduce, Spark, and Flink). They found ease of use, learnability, language support, auto-configuration, and community support can make big data platforms more usable to data scientists.
Mehta et al. \cite{mehta2016comparative} evaluated five large-scale image analysis systems (SciDB, Myria, Spark, Dask, and TensorFlow) and found various usability problems, including lack of support for user-provided Python code 
and manual tuning requirements for efficient execution.
These studies show that data science tools often fail to support data practitioners in certain data processing and analysis tasks beyond programming.
Since applying DP involves data science tasks, our study investigates if DP tools share similar usability issues as other data science tools.

Software engineering researchers have identified many usability issues in the technical documentation provided by developer- or programmer-facing software tools~\cite{treude2014extracting,meng2018application,aghajani2019software, aghajani2020software}, such as inconsistent content quality (e.g., readability, completeness, up-to-dateness) and poor navigation within the documentation.
Additionally, Becker et al.'s systematic review of text-based programming error message research revealed diagnostic messages generated by compilers are often unhelpful to programmers~\cite{becker2019compiler}. 
For example, compiler error messages written in natural language were as difficult to read as source code~\cite{barik2017}, and many error messages were poorly designed, particularly for novice programmers~\cite{prather2017}. 
Recommendations to improve programming error messages include increasing error message readability, reducing users' cognitive load, providing context to localize the problem, and showing examples, solutions, or hints to programmers~\cite{traver2010compiler,becker2016,barik2017,Hartmann2010}. 
In our study, we also examine how DP tools could leverage existing usability best practices from these non-DP software tools.
\section{Methods and Study Design}
We chose usability testing methods~\cite{nielsen1994usability, dumas1999practical} to observe data practitioners' efforts to understand and implement DP using DP tools. Usability testing can identify impediments to data practitioners' DP implementation. 
We used surveys, interviews, and think-aloud protocol~\cite{cooke2010assessing} for the data to answer our research questions. We executed the usability test remotely to reach a wider pool of participants. Research has shown that remote synchronous usability tests align closely in efficacy with traditional lab-based tests~\cite{usabilitytest07}. 

\subsection{Selection of Differential Privacy Tools}
\label{sec:select-diff-priv}
To select tools for our study, we first conducted a review of available DP tools and decided on inclusion criteria based on study goals and feasibility that would allow for direct comparisons between tools. We required that tools: (1) be open source, (2) support standard statistical queries (count, sum, average, etc.), (3) have comprehensive documentation, and (4) provide a Python API. Based on these criteria, we did not include graphical applications like DPCreator~\cite{dpcreator} or Private data Sharing Interface (PSI)~\cite{gaboardi2016psi}, or machine learning tools~\cite{papernot2019machine, yousefpour2021opacus}. We eliminated Chorus~\cite{chorusweb, johnson2020chorus}, GoogleDP~\cite{googledp}, Privacy on Beam~\cite{privacybeam}, PINQ~\cite{mcsherry2009privacy}, and ZetaSQL~\cite{zetasql, wilson2020differentially} due to lack of Python support. We included the remaining four tools in our study: OpenDP~\cite{opendp}, PipelineDP~\cite{pipelinedp}, DiffPrivLib~\cite{diffprivlib}, and Tumult Analytics~\cite{berghel2022tumult}.

\subsection{Study Procedures}
\label{sec:study_procedures}



\begin{table}
\footnotesize
\centering
\begin{tabular}{|cccccc|}
\hline
\textbf{ID} & \textbf{Tool} & \textbf{DP} & \textbf{DP Answers} & \textbf{Black or} & \textbf{Non-} \\
 & & \textbf{Expertise} & \textbf{Correct} & \textbf{Hispanic} & \textbf{Male} \\
 \hline
E005 & DiffPrivLib & Expert & 4/4 &  &  \\
E008 & DiffPrivLib & Expert & 3/4 & x & x \\
E011 & DiffPrivLib & Expert & 4/4&  & x \\
N002 & DiffPrivLib & Novice & 0/4&  & x \\
N004 & DiffPrivLib & \phantom{$^\dagger$}Novice$^\dagger$ & 3/4 &  & x \\
N011 & DiffPrivLib & Novice & 1/4 &  & x \\
\hline
E002 & OpenDP & Expert & 3/4 &  &  \\
E007 & OpenDP & Expert & 4/4 &  &  \\
E012 & OpenDP & Expert & 4/4&  &  \\
N003 & OpenDP & Novice & 2/4 & x & x \\
N008 & OpenDP & Novice & 2/4 &  & x \\
N012 & OpenDP & \phantom{$^\dagger$}Novice$^\dagger$ & 3/4 & x & x \\
\hline
E001 & PipelineDP & Expert & 4/4 &  & x \\
E004 & PipelineDP & Expert & 4/4 &  &  \\
E009 & PipelineDP & Expert & 3/4 &  &  \\
N005 & PipelineDP & Novice & 1/4 &  & x \\
N009 & PipelineDP & Novice & 1/4 &  & x \\
N013 & PipelineDP & Novice & 1/4 &  &  \\
\hline
E003 & Tumult & Expert  & 3/4 &  &  \\
E006 & Tumult & Expert & 4/4 & x &  \\
E010 & Tumult & Expert & 4/4&  &  \\
N006 & Tumult & Novice & 1/4 &  & x \\
N007 & Tumult & \phantom{$^\dagger$}Novice$^\dagger$ & 3/4 &  &  \\
N010 & Tumult & Novice & 0/4 &  & x \\
\hline
\end{tabular}
\caption{Summary of 24 study participants. The $\dagger$ symbol denotes participants who were initially categorized as DP experts by the eligibility survey but then re-categorized based on incorrect answers to DP questions in the post-task interview. }
    \vspace*{-10pt}
\label{tbl:participants}
\end{table}

\subsubsection{Recruitment \& Screening}

This study received approval from our university's Institutional Review Board (IRB).
We conducted a pilot study with four graduate students (one per tool) from our university and compensated them 25 US dollars each. Outcomes included adjusting study time allocation, increasing participant compensation, and clarifying survey and interview questions. 

We aimed to recruit at least 24 \emph{data practitioners}, with a balanced ratio between \emph{DP novices} and \emph{DP experts}, according to best practices for usability testing with developers in the privacy and security field~\cite{acar2016you}. We posted the study recruitment advertisement with a link to our eligibility survey on the Women in Machine Learning and OpenDP mailing lists, on Reddit in data science-related subreddits, and on LinkedIn. 

The eligibility survey (Appendix~\ref{sec:eligibility-survey}) determined participants' eligibility, obtained potential participants' informed consent, and gathered information about their data science and DP expertise. We deemed respondents eligible if they self-reported adequate data science experience (questions 1-3) and correctly answered at least one Python question (questions 4-5).
We initially categorized respondents to be \emph{DP experts} if they correctly answered 3 out of 4 DP knowledge questions (questions 8-11), and \emph{DP novices} otherwise. 
We finalized the DP expert/novice categorization after each session by assessing participants' answers to DP questions in the post-task interview (Appendix~\ref{sec:post-task-interview}) since multiple-choice questions in the eligibility survey were subject to guessing.
This led to the re-categorization of 3 participants as DP novices (see Table~\ref{tbl:participants}). 

Of the 109 respondents who started our eligibility survey, 83 completed it and 47 were eligible. 
We invited all 47 eligible respondents to the study, prioritizing underrepresented females due to diversity goals and timeline constraints.

We chronologically assigned confirmed participants to tools equally using the initial DP expert/novice categorization. 
After the initial tool assignment, we confirmed with each participant that they had not used the assigned DP tool before.
We continued recruitment after adjusting 3 participants' expert/novice categorization until we reached our recruitment target with a balanced expert/novice ratio.

26 confirmed participants completed their study sessions but we excluded two from data analysis (N001, E012): One due to the participant's inadequate Python skills, and the other due to an unavoidable session disruption that shortened task completion time.  A summary of the 24 study participants, their tool assignments, and their responses to the eligibility survey appear in Table~\ref{tbl:participants}. 
Participants' ages spanned from 18 to 40, but most (14) fell between 25-34 years. Our sample consisted of 54\% females, 38\% males, 4\% non-binary individuals, and 4\% who chose not to specify their gender. We conducted all usability test sessions on Microsoft Teams, following specific guidelines to maintain consistency. After the study session,  each participant was compensated with a gift card of 40 US dollars for up to 1.5 hours of study time.



\subsubsection{Pre-task Procedures} Before commencing usability study tasks, we made sure participants shared their screens and understood the think-aloud protocol. Participants also reviewed a handout that covered the fundamentals of DP and a tutorial of their assigned DP tool with executable sample DP tasks in Jupyter Notebook (see Appendix~\ref{sec:handout}). 
The handout and the tutorial provided participants necessary background for the study tasks but may introduce confounding factors to study results (detailed in Section~\ref{discussion:limitations}). 

We informed participants that 
they could refer back to the handout and the tutorial, consult the tool's official documentation, and use Google search during the study. 
We asked them not to use how-to resources, like StackOverflow. This ensured that participants had access to essential general resources (e.g., Python libraries) to complete the study tasks, but not to existing solutions to prevent cheating. 

\subsubsection{Usability Testing Tasks}
We designed three usability testing tasks on differentially private data analysis, shown in Table~\ref{tab:usability_tasks}. We modeled the tasks on a demo in Pipeline DP's documentation~\cite{pipelinedp}, changing it to a new, synthetic dataset that counted restaurant visits across a week (see Appendix~\ref{app:data}). 
Our tasks involved common data analysis operations supported by all four DP tools (i.e., count, sum, mean). Participants had one hour to complete the tasks by writing Python code in a shared Jupyter notebook.  

\begin{table}
  \centering
  \begin{tabular}{|p{.05\textwidth} | p{.35\textwidth} |}
    \hline
    \textbf{Task} & \textbf{Description}\\
    \hline
    Task 1 & 
    How crowded is the restaurant on weekdays? (total number of visits for each weekday) \\
    \hline
    Task 2 &
    Total amount of time spent by visitors on each weekday (exclude weekends). \\
    \hline
    Task 3 &
    Average amount of time spent by visitors on each weekday (exclude weekends) \\
    \hline
  \end{tabular}
  \caption{Usability testing tasks. See Appendix~\ref{app:data} for details of the dataset used and Appendix~\ref{app:task-solutions} for solutions.}
      \vspace*{-10pt}
  \label{tab:usability_tasks}
\end{table}

The three tasks were the same across DP tools. The assigned total privacy budget was $\epsilon = 1.2$ for all the tasks. Participants could set all other parameters themselves (including the per-task privacy budget). We encouraged participants to articulate their thought process using the think-aloud method, and we recorded both their spoken insights and on-screen actions during the study.


\subsubsection{Post-task Procedures} Participants completed a post-task survey and a post-task interview (Appendices \ref{sec:post-task-survey} and \ref{sec:post-task-interview}). The survey repeated DP questions from the eligibility survey to assess participants' DP understanding after the study. It also gathered data on participants’ study experiences. The post-task interview gathered qualitative data for deep insights into participants’ challenges during the study and their preferences for DP tools.

\subsection{Usability Measurements and Data Analysis}
\subsubsection{RQ1: DP Understanding}
Even experienced data scientists sometimes fail to grasp the intricacies of DP~\cite{cummings2021need, xiong2022using}. The DP tools in our study all aim to make DP more understandable to data practitioners.
To assess these tools' effectiveness in supporting DP understanding, we used the same four DP knowledge questions in the eligibility survey and the post-task survey to compare participants' pre-task and post-task DP knowledge differences. To mitigate confounding factors introduced by pre-task procedures, we also analyzed participants' explanations of key DP concepts in our post-task interview and their reported useful sources for DP understanding from post-task survey and interview.

\subsubsection{RQ2: DP Implementation} 
We used three widely-used usability criteria --- learnability, efficiency, and error prevention --- to assess how effective the tools support DP implementation~\cite{nielsen1996usability}.
\textbf{Learnability} measures if new users can successfully use a specific tool or interface. We use task success and failure rates~\cite{bevan2006practical} to measure DP tools' general learnability. Specifically, we evaluated whether users succeeded or failed to complete tasks and assessed the correctness of their completed tasks against our reference solutions.
\textbf{Efficiency} measures how fast users can accomplish tasks with a specific tool or interface. We recorded the time taken to complete each task to measure DP tools' efficiency. 
\textbf{Error prevention} is about how well a tool prevents user errors and, in the cases of error, how well a tool facilitates error identification and recovery.
We define errors during DP implementation as interruptions of progress toward task completion and qualitatively analyzed these interruptions from the screen recordings, think-aloud, and post-task interviews.
Additionally, we analyzed participants’ post-task survey responses to identify the factors that impacted DP implementation.

\subsubsection{RQ3: User Satisfaction} 
We first quantitatively evaluate user satisfaction using the two standardized measurements: the System Usability Scale (SUS) ~\cite{brooke1996sus} and the Net Promoter Score (NPS) ~\cite{grisaffe2007questions}. 
Since DP tools are specialized data science tools, we slightly customized the wording of SUS and NPS.
We also analyze the qualitative data from post-task interviews, including their overall user experiences and areas of improvement, to yield insights into user satisfaction with these DP tools.

\subsubsection{Mixed-Methods Data Analysis}
We report the descriptive statistics by tool to allow usability comparison across the four DP tools examined. We also report key statistics by participants' prior DP expertise level, either expert or novice, so that usability is recognized relative to participants' prior DP knowledge. Due to the small sample size, we refrained from performing statistical tests to avoid over-generalization (details in  Section~\ref{discussion:limitations}). 

The first and the second authors also rigorously analyzed the qualitative data collected from this study, including transcripts of audio recordings, video recordings of participants' screens, and Jupyter notebooks from all sessions. 
The two authors used a hybrid thematic analysis approach combining inductive and deductive coding~\cite{fereday2006demonstrating} to annotate the data.
They created the initial codebook from the pilot sessions and continuously refined it through research team discussions during the full study data analysis. The finalized codebook included both qualitative codes (e.g., type of challenges during implementation, misunderstandings of DP concepts) and quantitative counts derived from qualitative assessment (e.g., number of correctly completed tasks, time taken for each task). 
Then the first and the second authors independently coded all qualitative data using the codebook. They resolved all coding conflicts either through their own discussion or through seeking consensus from the whole research team.


\section{Results}

\subsection{RQ1: DP Understanding} \label{results:RQ1}

\begin{figure*}
    \centering
    \begin{subfigure}[t]{0.38\textwidth}
        \centering 
        \includegraphics[width=.8\textwidth]{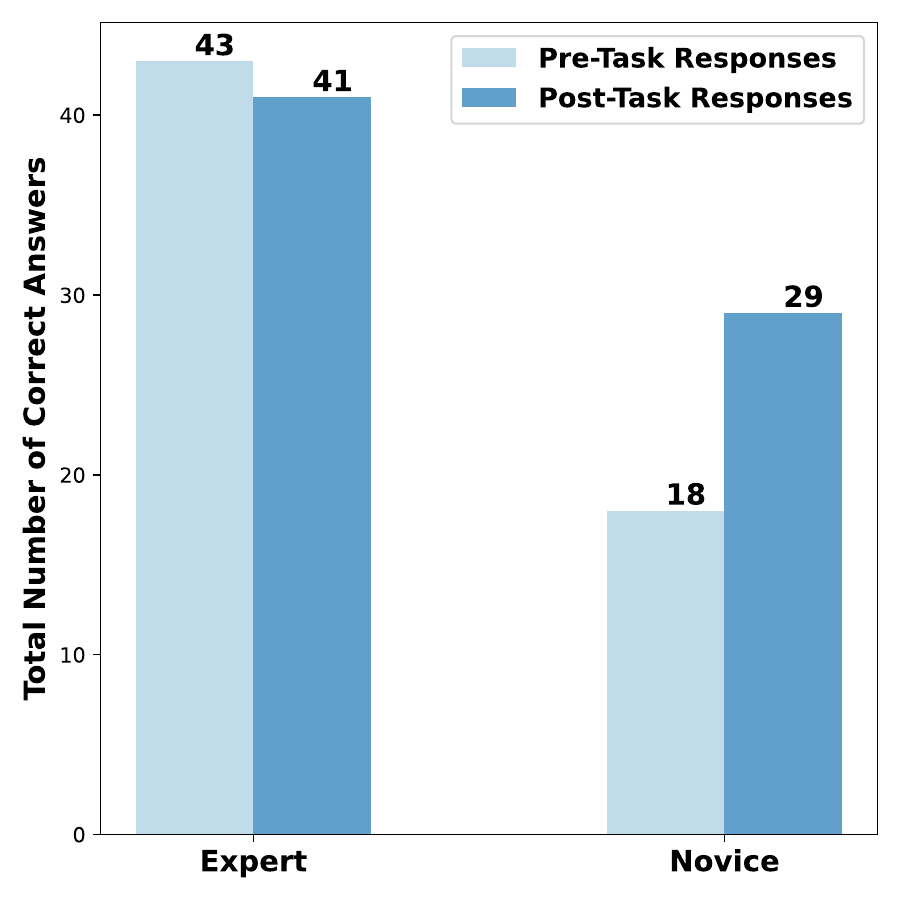}%
        \caption{By expertise level}
        \label{subfig:expertise}
    \hspace*{-20pt}
    \end{subfigure}%
    ~
    \begin{subfigure}[t]{0.64\textwidth}
        \centering
        \includegraphics[width=.8\textwidth]{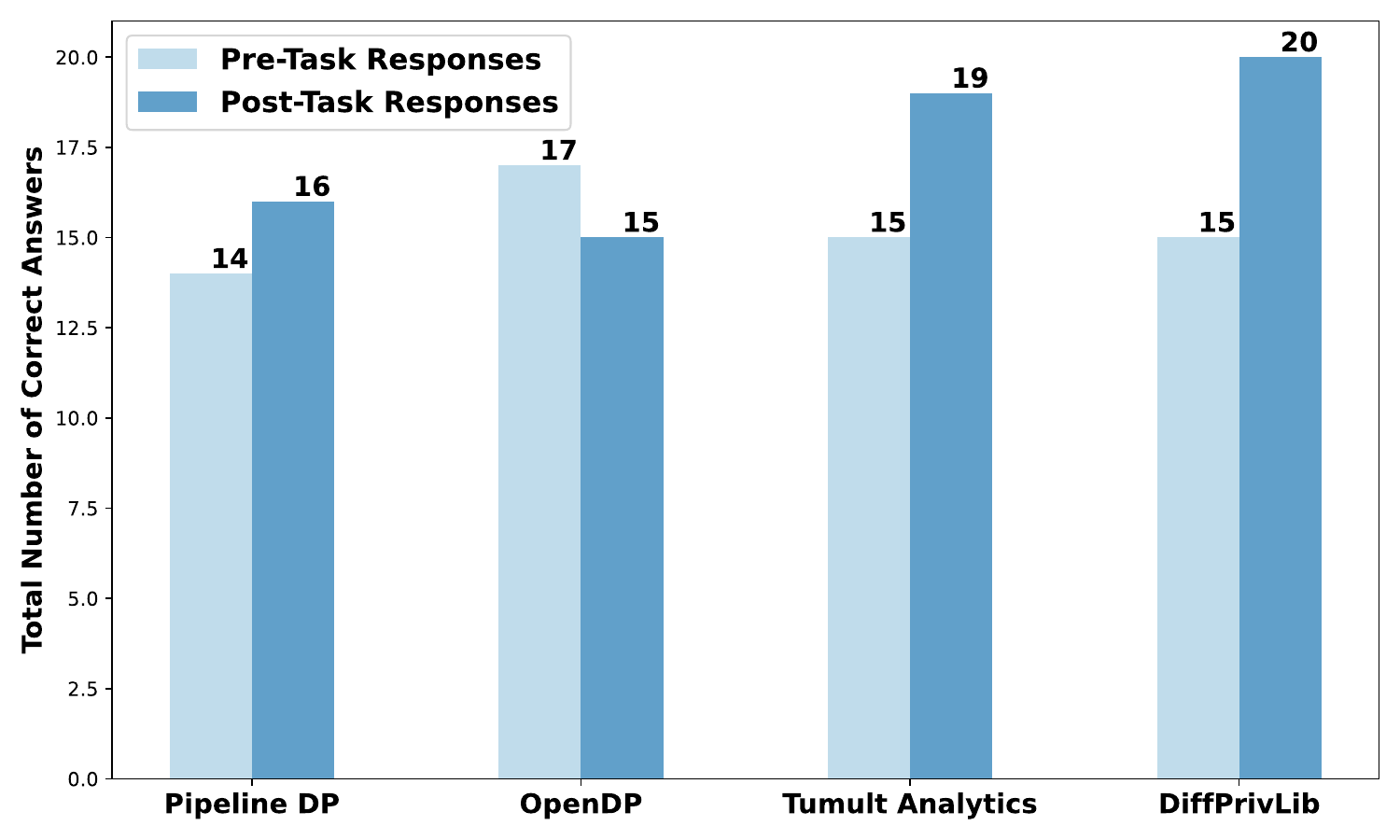}
        \caption{By tool} 
        \label{subfig:tools}
    \end{subfigure}%
    \vspace*{-10pt}
    \caption{Total number of correct answers to DP knowledge questions before and after study tasks.}
    \vspace*{-10pt}
    \label{fig:pre_post_results}
\end{figure*}



\subsubsection{Pre- and Post-Task Response to DP Questions}\label{sec:pre-post-DP}
Figure~\ref{fig:pre_post_results} reports the number of correct answers to the DP knowledge questions before and after study tasks, organized by participants’ DP expertise level and by tool.
Specifically, Figure~\ref{subfig:expertise} shows that experts provided similarly high-level of correct answers pre- and post- tasks, possibly due to their familiarity with DP concepts. However, novices showed a boost in their DP understanding as shown by the rise in correct answers from pre-task to post-task. 
This result indicates that our study procedures, including the DP implementation tasks, particularly helped novices understand DP concepts. 
Figure~\ref{subfig:tools} shows the pre- and post-task DP knowledge difference across tools.
All of the tools except OpenDP boosted participants' DP understanding, where DiffPrivLib saw the greatest jump from 15 to 20 correct answers.
Note that the study-provided handout and tutorials also impact participants' post-task DP understanding (see Section ~\ref{sec:RQ1:useful_source}). 
                                     
\subsubsection{Participants' Explanation of DP Concepts}
\label{sec:part-expl-dp}
To further investigate participants' understanding of DP, we looked at how they described key DP concepts in their own words during the post-task interview (Appendix~\ref{sec:post-task-interview} questions 2-3), focusing on DP, $\epsilon$, privacy budget for each task, and total privacy budget for all tasks. 
We aimed to see if participants understood that the privacy budget and $\epsilon$ essentially refer to the same concept and that the total privacy budget across multiple analyses accumulates the $\epsilon$ values (i.e., sequential composition). We considered participant responses to be correct if they were factual and includes details 
similar to our sample correct answers in Appendix~\ref{sec:post-task-interview}. 

Table~\ref{dp_concept_expertise} details the number of participants who could accurately explain DP concepts, divided both by their level of expertise and by the assigned tool. All 12 expert participants accurately explained the concept of DP. For example, one expert provided a robust definition, stating,  \textit{"Differential privacy is a mathematical definition for privacy that basically says that if we compute an analysis with a particular individual's data or without it, we should get similar outputs. Whether or not somebody participates in the data set, the outcome should be pretty much the same"}(E011). This explanation is correct because it clearly describes how randomness is used in analyses to ensure results are consistent, whether an individual's data is included or not, and emphasizes the importance of the privacy parameter. 
When discussing the privacy budget, 11 out of 12 experts explained how the budget was allocated in individual tasks, and 10 out of 12 were able to describe how these budgets add up to form the total privacy budget. 

Only 9 out of 12 novices could adequately explain DP in their own words, often missing critical details. A typical novice explanation was less precise, \textit{"From what I remember, it's like some sort of tool-based guarantee for privacy over millions of users..."} (N006), which lacks specificity and critical details about the mechanism of DP, like the privacy parameter. 
In their understanding of the privacy budget for each task, 8 out of 12 novices had a basic grasp. 7 out of 12 demonstrated an understanding of privacy budget accumulation, suggesting areas of confusion among novices.

The above difference between experts and novices shows the importance of participants' prior DP knowledge on their understanding. 
Additionally, each assigned tool had no clear impact on participants' understanding of DP, as reflected in Table ~\ref{dp_concept_expertise}.

Notably, participants gave incorrect answers for the question ``what was the total privacy budget for the whole notebook?'' more often than for the other questions (Table~\ref{dp_concept_expertise}). This result was consistent across experts and novices, and across tools, suggesting that composition is a difficult concept and should be made clear by DP tools. For example, one novice participant answered: \textit{"I got confused between these, total budget and the amount of epsilon for each of the individual tasks. That part, I didn't get it<"} (N010)

\begin{table*}
\small
  \centering
  \begin{tabular}{|p{.4\textwidth} ||c|c||c|c|c|c| }
    \hline
    \textbf{Question} & \textbf{Experts} & \textbf{Novices} & \textbf{DiffPrivLib} & \textbf{OpenDP}& \textbf{PipelineDP} & \textbf{Tumult}\\
    & n = 12 & n = 12 & n = 6 & n = 6 & n = 6 & n = 6 \\ 
    \hline
    After completing the tasks, can you explain differential privacy to me in your own words? &12 &9  &5 &6 &5 &5 \\ 
    \hline
    What was the privacy budget for each task? & 11 & 8 & 4 & 5 & 5 & 5\\
    \hline
    What was Epsilon? & 12 & 9 & 5 & 6 & 5 & 5 \\
    \hline
    What was the total privacy budget for the whole notebook? & 10 & 7& 4 &4 & 4 & 5 \\
    \hline
  \end{tabular}
  \caption{Number of correct answers to post-task survey questions measuring the understanding of DP concepts, disaggregated by level of expertise and by tool. Experts answered more of these questions correctly than novices, but the assigned tool had no clear impact on the number of correct answers. See Appendix~\ref{sec:post-task-survey} for sample correct answers.}
  \label{dp_concept_expertise}
\end{table*}

\subsubsection{Useful Sources for DP Understanding} \label{sec:RQ1:useful_source}

\begin{figure*}
    \centering
    \begin{subfigure}[b]{0.6\textwidth}
        \centering 
        \includegraphics[width=\textwidth]{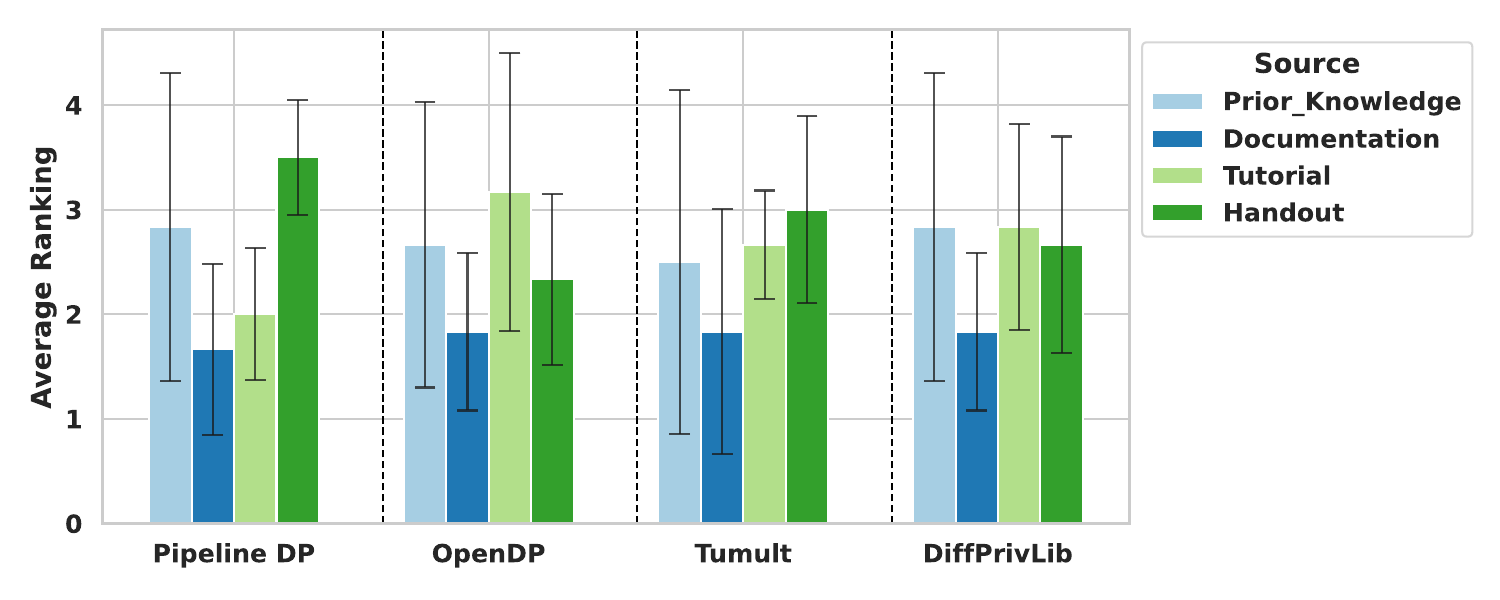}
        \caption{Useful sources by tool}
        \label{subfig:conceptual_understanding_tool}
    \end{subfigure}
    \hfill 
    \begin{subfigure}[b]{0.37\textwidth} 
        \centering
        \includegraphics[width=\textwidth]{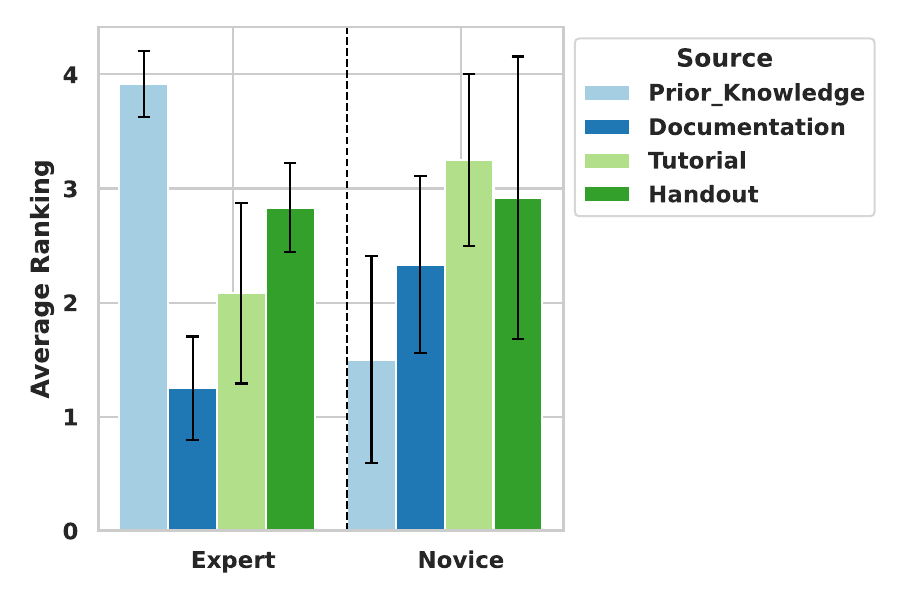}
        \caption{Useful sources by expertise level}
        \label{subfig:conceptual_understanding_expertise}
    \end{subfigure}%
    \caption{Useful sources that support participants' DP understanding, by tool (a) and by expertise level (b).}
    \label{fig:conceptual_understanding}
\end{figure*}

Participants selected all sources that helped them understand DP concepts during the study in the post-task survey, as shown in Figure~\ref{fig:conceptual_understanding}. The figure displays the average rankings of resources, where resources are ranked based on participants' preferences from 1 (most helpful) to 4 (least helpful).
Figure~\ref{subfig:conceptual_understanding_tool} indicates that the handout and the tutorials supported their DP understanding more than DP tools' official documentation across all tools, while participants' prior DP knowledge played a key role.
Figure~\ref{subfig:conceptual_understanding_expertise} shows that experts relied heavily on their prior DP knowledge, while novices used the handout and the tutorials to understand DP concepts.
This result suggests educational sources like the handout and the tutorials provided in this study benefit data practitioners' DP understanding, while DP tools' documentation lacks such support.


Post-task interview data suggests that concrete examples (like the ones in our tutorials) and short explainers (like the handout) helped participants understand important DP concepts, as E001 commented:~\textit{"It also helped to have the tutorial. ... if you had only given me the documentation... it would have taken me much longer to put it together."}
Note that the handout and the tutorial were part of the study instrument, so we cannot fully attribute novices’ increased DP understanding in Figure~\ref{subfig:conceptual_understanding_tool} to the DP tools themselves.

Moreover, our qualitative data indicates that participants could use more help with DP's actual privacy protection. In the case of $\epsilon$-values and privacy budgets, we asked participants how strong they thought the privacy protection was for their just-completed task. One DP expert (E006) confidently said:\textit{"That's the hard question to answer. The total privacy budget for all of the tasks was 1.2, a value that is in line with recommended guidelines. [$\epsilon$] is around 1.0. So, maybe that’s somewhat strong."}. 
Other responses lacked consistency and confidence: \textit{ "I think [$\epsilon$] should be much lower...probably around .5 or probably even lower..."}(E003) and \textit{"Pretty strong...very strong, actually"}(N007).


%

\subsection{RQ2: DP Implementation}


\begin{figure*}
    \centering
    \begin{subfigure}[b]{0.47\textwidth} 
        \centering
        \hspace*{-20pt}\includegraphics[width=1.2\textwidth]{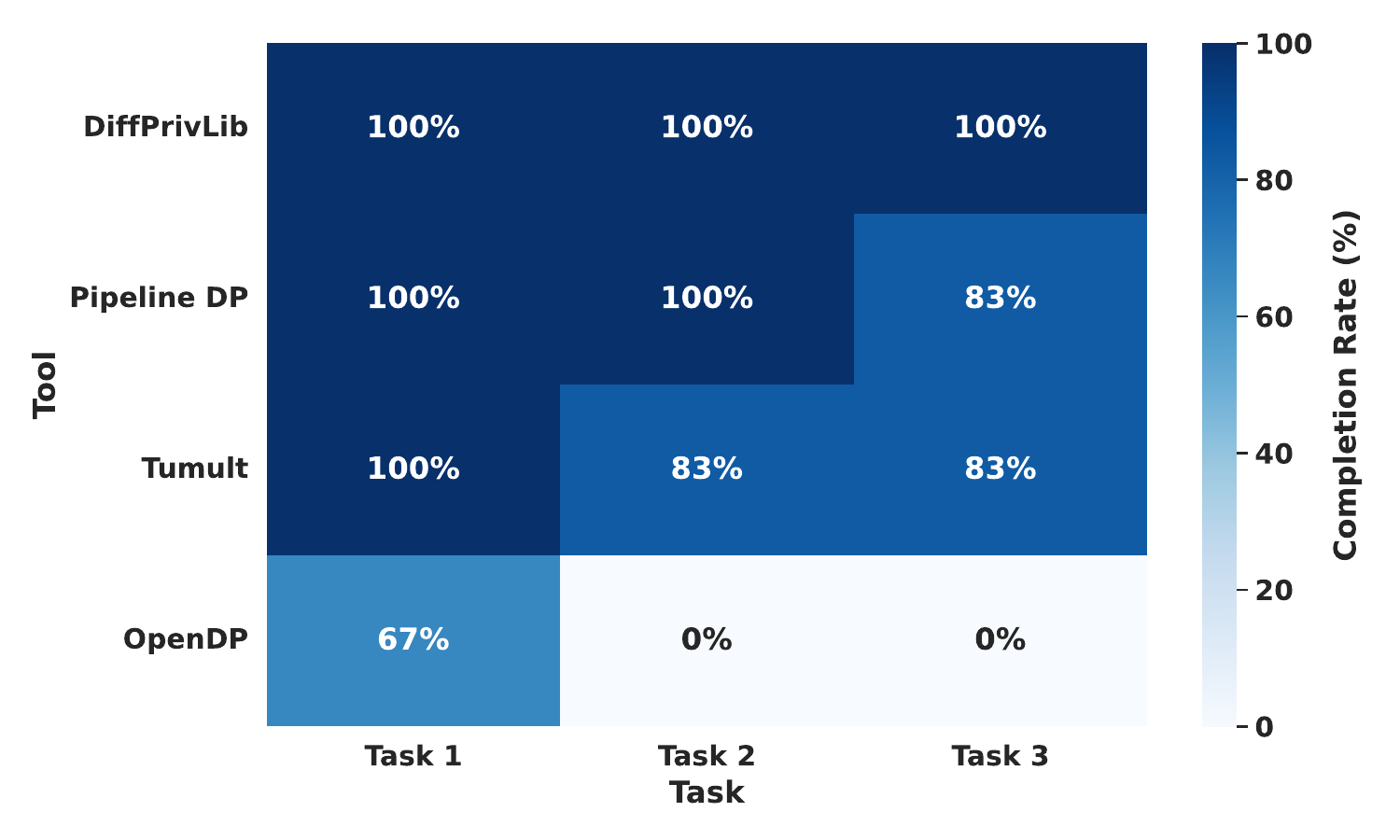}
        \caption{Task completion rates by task and tool} 
        \label{subfig:success_rate}
    \end{subfigure}%
    \hfill 
    \begin{subfigure}[b]{0.47\textwidth}
        \centering 
        \hspace*{-11pt}\includegraphics[width=1.2\textwidth]{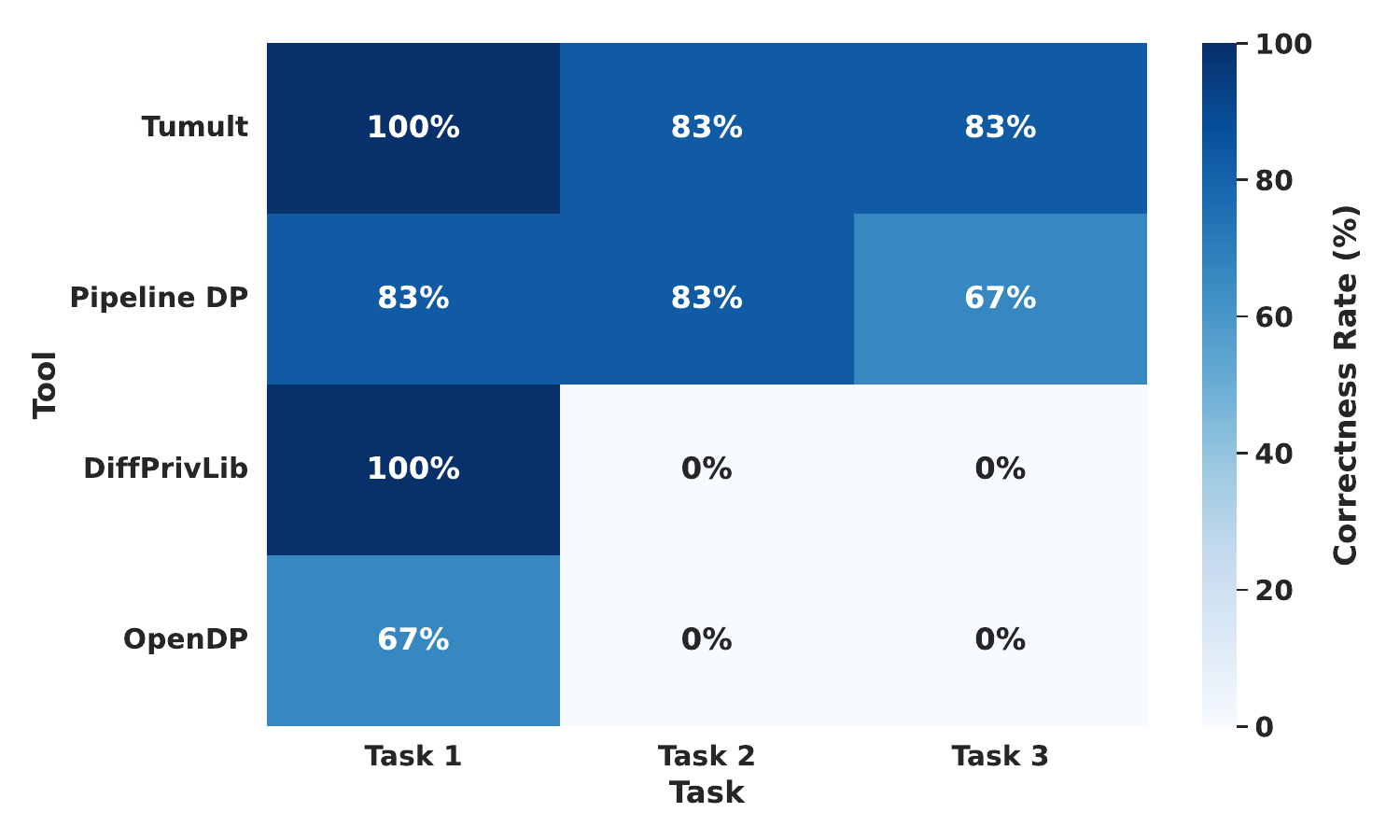}
        \caption{Task correctness rates by task and tool}
        \label{subfig:correctness_rate}
    \end{subfigure}
    \caption{Learnability of DP tools measured by (a) task completion rates and (b) task correctness rates. Each cell represents the percentage of participants who completed or correctly completed the task using the tool.}
    \label{fig:task_results}
\end{figure*}

\begin{figure*}
    \centering
    \begin{subfigure}[b]{0.6\textwidth}
        \centering 
        \includegraphics[width=\textwidth]{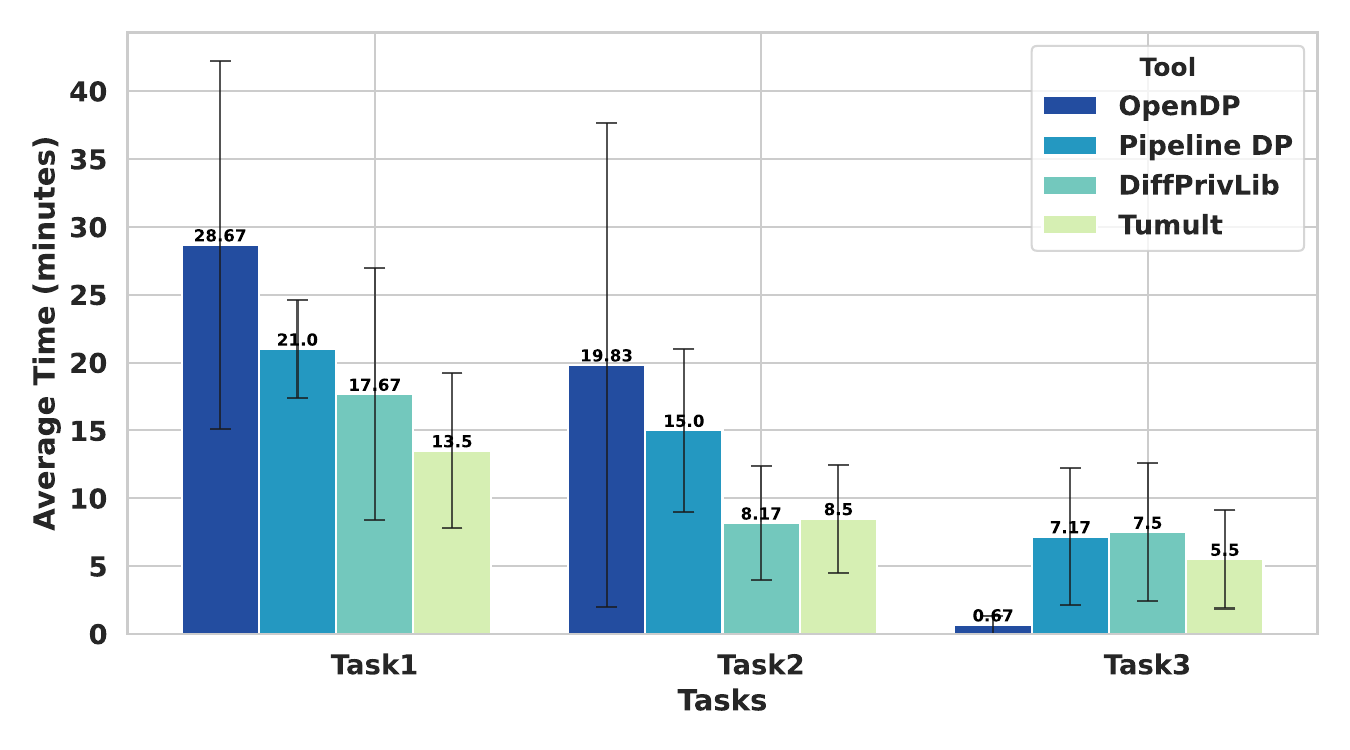}\vspace*{-6pt}
        \caption{Average time taken by tool}
        \label{subfig:avg_time_task}
    \end{subfigure}%
    \hfill 
    \begin{subfigure}[b]{0.4\textwidth} 
        \centering
        \includegraphics[width=\textwidth]{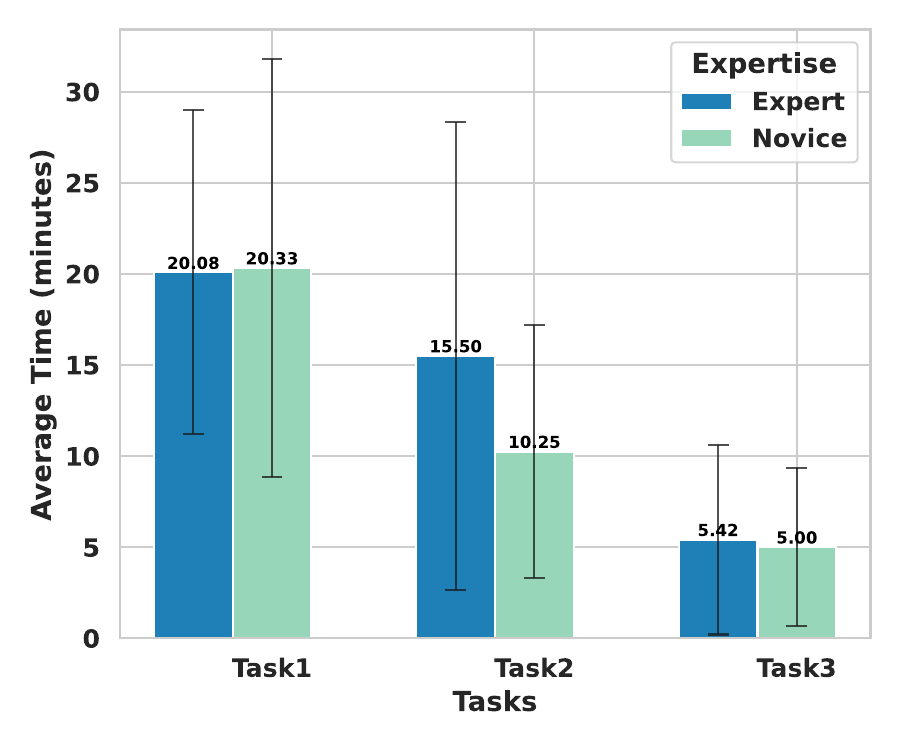}\vspace*{-6pt}
        \caption{Average time taken by expertise level} 
        \label{subfig:avg_time_expertise}
    \end{subfigure}%
    \caption{Average task completion time: (a) by tool (b) by expertise level.}
        \vspace*{-10pt}
    \label{fig:task_time_results}
\end{figure*}

\subsubsection{Learnability}  \label{results:RQ2:learnability}
\paragraph{Task completion and correct rates.}
To measure learnability, we evaluated the completeness and correctness of participants' solutions. We considered tasks \textbf{complete} when code executed without error and produced correctly formatted responses, and \textbf{correct} when they satisfied DP and had comparable utility to our reference solutions.

Figure~\ref{subfig:success_rate} shows the completion rates for three usability testing tasks across four tools: all DiffPrivLib participants completed all three tasks, while none of the OpenDP participants completed tasks \#2 or \#3. Tumult Analytics and PipelineDP results fall between these two extremes, with all participants completing at least task \#1.

The varying completion rates may derive from the different API designs of the tools. DiffPrivLib provides a minimal API and encourages users to use it in combination with Python data analytics libraries like Pandas. Similarly, Tumult Analytics mimics an existing data analytics API called Spark. OpenDP, in contrast, does not leverage mainstream Python libraries for a learning scaffold. 
Participant comments on API design from post-task interviews lend support to this finding. For example, one expert (E006) liked the similarity of the Tumult Analytics to Spark: \textit{"I think the fact that it was very similar to Spark was really helpful...I have a decent amount of experience with Spark and Pandas, so that was very intuitive."}
%

Figure~\ref{subfig:correctness_rate} shows the task correctness rates.
Some participants completed tasks but incorrectly, so the correctness rates are no larger than the corresponding completion rates. Combined, the completion and correctness rates show that: (1) complete Tumult Analytics and OpenDP solutions were all correct; (2) complete PipelineDP solutions were mostly---but not all---correct; (3) Complete DiffPrivLib solutions were all \emph{incorrect} for tasks \#2 and \#3.

\paragraph{Causes for incorrect implementation.}
Qualitative analysis of the screen recordings revealed the causes of some incorrectly completed tasks
First, all six DiffPrivLib participants failed to apply the correct \textbf{sensitivity}, which refers to the upper and lower bounds that provide the extent of valid DP, in tasks \#2 and \#3. (Task \#1, a counting query, has a sensitivity of one, a value that is intuitively correct.) DiffPrivLib does not signal any error related to sensitivity bounds, even though this mistake violates DP. Some expert participants were uneasy about their approach for setting sensitivity, but even these participants were not able to produce correct solutions.

Second, some tools lack \textbf{feedback} about query results' correctness. For example, one PipelineDP participant (E004) used strings (rather than integers) as grouping keys, resulting in histograms containing only 0s, and the participant did not notice the mistake.
The participant later discussed this in the post-task interview, \textit{"It's the right number of attributes and it's the right metric...the result is very noisy,"} but he added, \textit{"I don't know if there's a way to check the final [privacy] budget."}
In this case, Pipeline DP's lack of feedback affected the solution's correctness but did not violate DP.

Finally, confusion about \textbf{whether and how the tools handle the privacy budget} led to incorrectness, particularly for Pipeline DP and DiffPrivLib. E009 commented on PipelineDP:
\textit{"I would expect maybe that [a] budget accountant object could tell me my budget so far. [I'm] looking for a way to figure out how much I spent so far."}
And N011 on DiffPrivLib:
\textit{"[I'm] confused about how the privacy budget would be handled at the object level.
When creating the mechanism objects, should I use the same object for every analysis...and the $\epsilon$ will add up to the right number...can you compose all of those together? That wasn't totally clear to me."}

\subsubsection{Efficiency} \label{results:RQ2:efficiency}
To measure efficiency, we calculated the time taken to complete each task by reviewing the screen recordings. 

Figure~\ref{subfig:avg_time_task} shows the time taken on each task by tools. OpenDP participants spent the most time on task \#1 (nearly 30 minutes on average), while Tumult Analytics participants spent the least (fewer than 15 minutes on average), with DiffPrivLib and PipelineDP falling in between.
The time taken for Task \#2 shares a similar trend while all participants spent less time on task \#2 than task \#1. 
However, time taken for task \#3 varied. OpenDP participants spent almost no time on task \#3, while participants using the other three tools spent similar amounts of time on task \#3, but less than that of tasks \#1 and \#2. 
The total time limit (1 hour) imposed on all tasks may affect the time spent on task \#3. OpenDP participants spent nearly all of the allotted time on tasks \#1 and \#2, leaving little time for task \#3. Participants using the other tools either finished task \#3 quickly or ran out of time. \textit{"I think I wasted a lot of time trying to find what I don't know,"} said E013.

Figure~\ref{subfig:avg_time_expertise} shows the time spent on each task, by participants' expertise level. For tasks \#1 and \#3, novices and experts took roughly the same amount of time; for task \#2, however, experts took \emph{longer} than novices. 
Qualitative analysis from participants' think-aloud showed that experts' confidence, curiosity, and skills prompted them to explore task solutions. E005 spent time \textit{"investigating the number of visitors that show up multiple times per day"} only to find the occurrences are rare in the data, concluding that \textit{"we can just set the sensitivity to one."} Other experts also spent time examining API functions, honing DP parameters, checking results, and exploring alternative approaches.
Novice users, in contrast, typically accepted the tool's default settings and did not spend time considering these issues. \textit{"I'm not familiar with all the different functions,"} N011 told us in think-aloud while looking at different options for DP noise, and added post-task, \textit{"The land of functions are [sic] totally wild to me."}

\subsubsection{Error prevention} \label{results:RQ2:error}

We consider interruptions of progress toward task completion as errors during DP implementation and call them "stucks". 
We also examine error recovery when participants resolve these interruptions and call them "unstucks."


\paragraph{Stuck and unstuck statistics.}
We report the counts for stuck and unstuck, as well as the unstuck percentages in Table~\ref{tab:stuck_unstuck}, organized by DP tool, by participants' DP expertise, and by stuck type (defined in Table~\ref{tab:stuck_type}).
Tool-wise, participants assigned to Tumult Analytics (38/43, 88\% ), DiffPrivLib (27/31, 87\%), and PipelineDP (38/54, 70\%) often managed to get unstuck, but those assigned to OpenDP rarely got unstuck (22/79, 28\%). Expertise-wise, DP experts (67/105, 64\%) and novices (58/102, 57\%) had similar unstuck percentages.

\begin{table}
  \centering
  \begin{tabular}{|p{.15\textwidth} | c | c | c |}
    \hline 
    \textbf{Tool} & \textbf{Stucks} & \textbf{Unstucks} & \textbf{Unstuck \%}\\
    \hline
    DiffPrivLib & 31 & 27 & 87.1\% \\
    OpenDP & 79 & 22 & 27.8\% \\
    Pipeline DP & 54 & 38 & 70.4\% \\ 
    Tumult & 43 & 38 & 88.4\% \\ 
    \hline
    \hline
    \textbf{DP Expertise} & \textbf{Stucks} & \textbf{Unstucks} & \textbf{Unstuck \%}\\
    \hline
    Experts & 105 & 64 & 63.8\% \\
    Novices & 43 & 16 & 56.9\% \\
    \hline \hline
    \textbf{Stuck Type} & \textbf{Stucks} & \textbf{Unstucks} & \textbf{Unstuck \%}\\
    \hline
    DP & 4 & 3 & 75.0\% \\
    Documentation & 65 & 31 & 47.7\% \\
    Python & 28 & 27 & 96.4\% \\ 
    Results & 18 & 4 & 22.2\% \\ 
    Task & 34 & 31 & 91.2\% \\ 
    Tool & 58 & 29 & 50.0\% \\ 
    \hline
  \end{tabular}
  \caption{Stuck counts, unstuck counts, and unstuck percentages  by DP tool, participants' DP expertise, and stuck type}
      \vspace*{-10pt}
  \label{tab:stuck_unstuck}
\end{table}

\paragraph{Stuck types.}
We identified six types of stuck in Table~\ref{tab:stuck_type} and contextualized them with qualitative data.
The most frequent stuck type was \textbf{documentation stucks} (65 counts, 48\% unstuck percentage), where participants had difficulty finding answers in tools’ documentation. \textit{"I can imagine how to do this without this library,"} said E011 (DiffPrivLib), \textit{"I'm trying to see...how to translate that into the library."} 
Second was \textbf{tool stucks} (58 counts, 50\% unstuck percentage), where participants struggled to execute tools’ function calls or to interpret tools’ error messages. Participants would either fail to grasp tool basics: \textit{"I don't get the terminology or the syntax,"} said E003 (Tumult Analytics); or, the issue with the tool was a specific aspect of DP: \textit{"I'm trying to figure out how I actually tell the session what the sensitivity of the query is,"} said E006 (Tumult Analytics). 
\textbf{Task stucks} were common (34 counts, 91\% unstuck percentage) but most participants got unstuck by asking researchers for task clarification. For example, days in our dataset are integers, 1-7. E005 (DiffPrivLib) asked, \textit{"Can I ask is day one equal to Monday and Day 7 equal to Sunday?"} 
Participants also experienced \textbf{Python stucks} (28 counts, 96\% unstuck percentage) but almost always got unstuck by consulting Python sources. 

\begin{table*}
  \centering
  \begin{tabular}{|p{.33\textwidth} | p{.66\textwidth} |}
    \hline
    \textbf{Stuck Type (Abbreviation)} & \textbf{Definition}\\
    \hline
    DP misunderstanding (DP) &
    Incorrectly interpreting or applying DP. \\
    \hline
    Documentation stuck (Documentation) &
    Struggle to interpret documentation descriptions. \\
    \hline
    Expected result stuck (Result) &
    Answer from a DP tool query that is not in line with expected DP values. \\
    \hline
    Python stuck (Python) &
    Don't know the correct Python or Pandas function to use. \\
    \hline
    Question stuck (Task) &
    Misinterpretation of a Task assignment, or need to clarify a Task detail. \\
    \hline
    Tool stuck (Tool) &
    Don't know the correct DP tool function to use. Failing to interpret error codes. \\
    \hline
  \end{tabular}
  \caption{Definitions of six types of stuck from our qualitative analysis}
  \label{tab:stuck_type}
\end{table*}

\paragraph{Usability issues.}
Our qualitative analysis articulated how DP tools' usability issues with their documentation and APIs caused errors and hindered error recovery. 
\textbf{DP tools' documentation} presented many usability problems. E001 found the upper bound for data values in PipelineDP unclear: \textit{ "I'm not super sure about this maximum value because I'm not sure if I interpret it correctly [in] the documentation."} 
Other participants hoped the documentation would provide more details about different API functions, such as the best DP mechanism for a data analysis task. E005 commented on DiffPrivLib's documentation, \textit{"I do think that sometimes when you present people with a suite of 16 options, it's important to detail what the differences are and when one option might be more effective than another."} 
The format of the documentation was also challenging. Participants struggled due to the lack of organization of OpenDP's documentation. E013, for one, \textit{"got lost in it."} 
\textbf{DP tools' APIs} also caused DP-specific errors. 
Participants struggled with API instructions to set parameters for DP mechanisms. And if the tools' parameters were idiosyncratic, the user interaction was less intuitive. \textit{"I was not confident because I didn't know what the library was doing"} and \textit{"I wasn't sure what the argument [meant],"} E004 (PipelineDP) said, \textit{"I don't really know in the end if I computed what I was really expecting to compute."}

Error recovery was challenging. N012 was frustrated by the lack of examples \textit{"...I couldn't get examples of people running into the same problem."} 
Error messages sometimes were unhelpful. For example, OpenDP's API returned error messages in Rust, and not translated to the API's Python wrapper. E002 said: \textit{"I don't really know any Rust. Coming from a Python experience, [it] might be better to have error messages in Python that indicate the error in the line of Python."}



\subsubsection{Factors impacting DP Implementation} \label{RQ2:factors}

Participants’ post-task survey responses revealed the factors that helped or hindered their DP implementation in the study. Full results appear in Figures~\ref{fig:factors_helping} and~\ref{fig:factors_hindering} in Appendix~\ref{sec:additional-results}.

9 out of 12 novices and all 12 experts reported the tutorial helped their DP implementation, with tool documentation (5 novices and 8 experts) and their data science skills (7 novices and 8 experts) close behind. 
Notably, none of the participants assigned to OpenDP reported that their data science skills or the tool's documentation were helpful, possibly due to how OpenDP's API differs from mainstream Python libraries. 
E002, someone familiar with data frames and method chaining in other Python libraries, failed to understand basic OpenDP syntax, \textit{"I don't know what you call that little stream operator thingy."} E012 said that \textit{"it's written in a very C-heavy style as opposed to a Python-style that most people are used to."}

8 out of 12 novices were hindered by lack of prior DP knowledge, while 5 out of 12 experts reported being hindered by DP tools' documentation in completing the tasks.
Novices like N009 (Pipeline DP) \textit{"took a lot of time understanding what the metrics are and what each parameter is"} and N010 (Tumult Analytics) \textit{"got confused between the total [privacy] budget and the [epsilon] for each of the individual tasks."} One expert, E003, asked for \textit{"a step-by-step guide on how you can how you can use Tumult Analytics for your particular use case."}
These results suggest that DP tools should help enhance novices' DP knowledge and improve their documentation to support experts' DP implementation.

\subsection{RQ3: User Satisfaction} \label{results:RQ3}
%
\subsubsection{Quantitative Ratings} \label{results:RQ3:quan}
The Net Promoter Score (NPS) and System Usability Scale (SUS) metrics from the post-task survey showed that participants were most satisfied with DiffPrivLib and least satisfied with OpenDP.
DiffPrivLib had the highest NPS (33.33), followed by Tumult Analytics (-16.67), PipelineDP (-33.33), and OpenDP(-66.67).
Similarly, DiffPrivLib had the highest SUS score (63.89), followed by Tumult Analytics (57.64), PipelineDP (54.51), and OpenDP (38.19). Full statistics appear in Figure~\ref{fig:satisfaction} in Appendix~\ref{sec:additional-results}. 
These ratings align with the task completion rates associated with each tool (Figure~\ref{subfig:success_rate})---DiffPrivLib had the highest user satisfaction ratings and the highest completion rate, followed by Tumult Analytics, PipelineDP, and OpenDP. This suggests that participants were most satisfied with tools that made it easy for them to complete the study tasks.


\subsubsection{Qualitative Results by Tool} \label{results:RQ3:qual}
Our qualitative results from the post-task interviews triangulated the above quantitative ratings and articulated participants' user experience with each tool, as described below.

\paragraph{DiffPrivLib} received positive comments about its API and documentation: \textit{"I liked the API of the tool. I thought the documentation was pretty clear" (E005)}.
Participants also liked its compatibility with familiar libraries: \textit{"I really liked that it integrated nicely into a library that I already have worked with, Pandas...\remove{acting as a layer on top of what I would already do.}" (E011)} and felt comfortable with the tool by the end of the session: \textit{"Now...I'm on task three, I feel like I have a hang of the pattern...this isn't adding that much more time to my typical process" (E011)}.
 
\paragraph{Tumult Analytics} was acclaimed for its intuitive API, as E010 said:\textit{"Similarity with Pandas was definitely a plus. That's probably the best thing they've done there.}" 
However, E003 expressed frustration with its documentation: \textit{"It was just a single-page documentation and I had to like scroll all the way down to find the exact syntax."}. 
The feedback addressed the user need for improved documentation navigation.

\paragraph{PipelineDP} exhibited documentation challenges:
\textit{"The documentation was quite incomplete...sometimes it just had one sentence about terms like 'Max contribution' or 'Max value' and it wasn't really clear to me what that meant" (E004)}.
Participants also wanted the ability to search:
\textit{"What [does] the documentation say about the budget? I don't have a way to search this page" (E001)} and found error messages confusing: \textit{"I think the error message wasn't super clear and it would be tough to debug" (E004)}. Several participants wished for examples in the documentation:
\textit{"Functions should contain some examples...[like] what each parameter is...\remove{ For somebody who is completely new...it is...difficult to understand}" (N009)}.

\paragraph{OpenDP} had usability issues with its error messages and documentation: \textit{"The error messages I'm getting here come from Rust and I don't know what it means" (E007)}, and
\textit{"The documentation wasn't useful...[I] felt like it was a little confusing...\remove{like a little cluttered...there's a lot of information}." (N008)}. OpenDP participants also wished for examples:\textit{"It would have been a lot more helpful if there were examples" (N012)}.

Overall, these findings on user satisfaction echo prior results --- DP tools' API design and documentation quality are paramount to data practitioners' user experience.


\section{Discussion and Recommendations}
\label{discussion}

\subsection{Limitations}
\label{discussion:limitations}
We acknowledge several limitations of this study. 
First, we only evaluated four open-source Python-based DP tools for fair comparison across tools so that our results cannot represent all DP tools. 
Similarly, the findings may not generalize to all data practitioners due to our small US sample. However, our sample is similar to prior usability studies evaluating security/privacy tools with developers~\cite{li2018coconut,naiakshina2017developers} to generate valid insights.
Therefore, we refrained from performing statistical tests to avoid over-generalization of the statistical results from this study to all DP tools or data practitioners. Instead, we emphasized key descriptive statistics and qualitative results.

Second, our study instrument introduced confounding factors because the handout and tutorials (Section~\ref{sec:study_procedures}) helped participants understand DP and complete study tasks. However, we had to prioritize study feasibility to ensure participants with varying prior knowledge had the necessary information to get started. To minimize this bias, we ensured our handout and tutorials did not reveal answers to study questions or tasks, and we remained cognizant when analyzing and reporting study results.(see Sections~\ref{sec:RQ1:useful_source} and ~\ref{RQ2:factors}).

Moreover, we only evaluated the usability of three first-step DP data analysis tasks. The results may not reflect the usability of the full capability of the examined DP tools. 
However, usability issues surfaced in these first-step tasks hinder developers' adoption of software tools or APIs~\cite{acar2016you,myllarniemi2018development}, our recommendations for usability improvements still benefit other DP tools and encourage overall DP adoption.

\subsection{Provide Usable Documentation}
Our results 
highlighted usability issues with DP tools' official documentation, leading to the following recommendations.

\paragraph{Improve documentation navigation.}
Participants generally experienced difficulty navigating DP tools' official documentation, including technical documentation on APIs.
Firstly, despite the fact that all four tools provide how-to guides with code examples in their documentation, participants struggled to find specific guides that matched their data analysis tasks at hand. For example, the descriptions of these guides are often generic and contain DP terminology (e.g., "how to perform counting queries with the Laplace mechanism"), which is unfriendly to DP novices.
The mismatch between documentation contents and the practical development tasks often caused poor documentation findability~\cite{aghajani2019software}, which can be mitigated by providing more accurate and readable task descriptions that align with users' goals~\cite{treude2014extracting}.
Secondly, our participants disliked DP tools' single-page formatting (see Section \ref{results:RQ3:qual}). This formatting uses a single web page to organize documentation for every API function within a module, which can be lengthy, worsening the findability problem. 
In contrast, mainstream Python libraries (e.g., NumPy, Pandas) use one page per documented function and are easier to navigate.
Additionally, some participants also hoped to be able to search within DP tools' documentation, which also resonates the proposed techniques to improve the usefulness of software documentation~\cite{aghajani2020software}.
Our findings echo prior software engineering research on usable documentation~\cite{aghajani2019software,aghajani2020software}. 
We believe DP tools can leverage existing best practices for good software documentation, such as providing intuitive task descriptions, improving information organization, and adding a search or recommender tool to improve documentation navigation.

\paragraph{Include DP-specific examples and advice.}
Many participants found the documentation for the API function they wanted to use but had trouble understanding the descriptions of DP-specific parameters and were not able to find examples that made use of the documented function.
Some requested more use cases and code examples within DP tools' documentation (see Section \ref{results:RQ2:error}). 
These results are consistent with prior research on developers' need for documentation~\cite{meng2018application,aghajani2020software}.
Moreover, some DP novice participants had trouble deciding which DP mechanism to use---for example, when given a choice between the Laplace, Gaussian, or Geometric mechanisms. Existing tool documentation fails to address these questions since it predominantly emphasizes \textbf{how} to use a specific mechanism rather than \textbf{which} mechanism to choose. To make DP tools' documentation truly usable for data practitioners, we recommend that DP tools go beyond generic best practices for usable documentation and include DP-specific advice that would particularly benefit DP novices.

\subsection{Improve Error Prevention \& Recovery}
The study findings yielded rich insights into how DP tools can prevent errors and help users recover from errors.

\paragraph{Warn users about severe DP violations.}
PipelineDP, Tumult Analytics, and OpenDP actively prevent DP violations---they require users to wrap sensitive data using special objects. They provide error messages when users attempt to perform actions that would violate DP. DiffPrivLib, on the other hand, relies on the user to avoid DP violations; for example, DiffPrivLib asks users to set the sensitivity for every mechanism and does not check that the specified sensitivity has been correctly enforced for the input data. This explains that all of the participants assigned to DiffPrivLib completed all three tasks, but \emph{every single participant} violated DP in their solutions for tasks \#2 and \#3 and failed to correctly complete them (see Figure~\ref{fig:task_results} in Section~\ref{results:RQ2:learnability}). 
Thus, we recommend that DP tools proactively warn users when DP might be violated.\footnote{DiffPrivLib raises a "privacy leakage warning" in some situations that may violate DP (e.g., when setting parameters based on the data), but not in all such cases. In particular, when the programmer uses an external library like Pandas to produce an aggregate result—as all of the participants in our study did—DiffPrivLib cannot enforce sensitivity bounds on the query and does not raise a warning.}

\paragraph{Improve error messages}.
When errors occurred, many participants had difficulty diagnosing and recovering due to poorly designed error messages (Sections \ref{results:RQ2:error} and \ref{results:RQ3:qual}). 
In particular, participants assigned to PipelineDP and OpenDP described confusion over the meaning of error messages, and trouble finding documentation to understand and fix the problem. This resonates with prior research on unhelpful compiler error messages of non-DP tools~\cite{barik2017, becker2016, prather2017}.
%
Additionally, OpenDP further confused users who primarily have a Python background with error messages generated in the programming language Rust.
We first recommend DP tools learning from general best practices to improve error message readability and provide examples, solutions, and hints~\cite{becker2019compiler}.
Additionally, DP tools should consider the average DP knowledge of their intended users and offer support when the error is DP-related (e.g., pointers to resources on DP violations).
%

\paragraph{Ensure clarity in privacy budget setting and tracking.} 
Some participants failed to explain the total budget (Section~\ref{sec:part-expl-dp}) and many were concerned with setting or tracking the privacy budget with different DP tools.
Tumult Analytics asks users to set the total and per-query budget with required API calls. 
This process was not as clear in other DP tools: Some participants assigned to PipelineDP and DiffPrivLib were not sure whether the library keeps track of the privacy budget at all. This confusion did not necessarily result in failure to complete the study tasks, but it would result in unintended DP violations in real-world implementations. 
We recommend that DP tools clearly convey how to set the privacy budget and how the tool accounts for the total budget.

\paragraph{Balance DP violation prevention and general usability.}
We also observe the tension between preventing DP violation errors and maintaining the tool's usability (Sections \ref{results:RQ2:learnability} and \ref{results:RQ3:quan}). OpenDP's strict API was effective at preventing DP violations, but OpenDP had lower completion rates and satisfaction ratings. DiffPrivLib's flexible API resulted in many DP violations but received high completion rates and satisfaction scores. 
Tumult Analytics seems to strike the best balance. Its API was effective at preventing DP violations where users had high completion rates and satisfaction ratings.
This indicates that DP tools may need to balance between their goal to prevent DP violation errors and the tool's usability.

\subsection{Make API Design Intuitive}
Our findings 
reveal participants' unique experiences with the APIs of DP tools, leading to the following recommendations.

\paragraph{Leverage users' familiarity with mainstream APIs.}
Results in Sections \ref{results:RQ2:learnability} and \ref{results:RQ2:efficiency} suggest that participants implemented DP more successfully with DP tools that incorporate mainstream APIs that they are familiar with.
Specifically, the intersection of DiffPrivLib with ubiquitous libraries like Pandas, garnered commendation. This cohesive integration provided a scaffold for new learning and obviated the need for relearning. 
Tumult Analytics was also appreciated for the way its API mimicked that of Spark. 
In contrast, PipelineDP provides an API centered on performing multiple aggregations at once, and OpenDP provides an API that focuses on transformations and composition. Neither is similar to mainstream data science APIs, which impeded participants' DP implementation in the study. 
To make APIs more usable, we recommend DP tools prioritize API designs that allow data practitioners to transpose their extant data science knowledge to the DP context, augmenting overall satisfaction.


\paragraph{Assist users in setting DP-related metadata via APIs.}
Setting DP-related metadata (e.g. total privacy budget, $\epsilon$ per query, upper bound on data values) is key to DP implementation.
DiffPrivLib includes \textbf{default values} for metadata. 
The choice to use default values simplifies the API, but may result in users accidentally accepting inappropriate default values. 
DiffPrivLib provides warnings when default values could result in DP violations. This helped participants to complete the tasks correctly and suggests that default values can be effective if appropriately selected and implemented.
The other three tools require \textbf{users to specify DP-related metadata} via APIs.
Experts appreciated that Tumult Analytics explicitly asks users to set per-query and total privacy budgets.
However, DP novices may struggle with manually setting metadata, like with other non-DP tools~\cite{mehta2016comparative}.
Participants found PipelineDP's API for setting metadata confusing and struggled with settings like \texttt{max\_value}, \texttt{partition\_extractor}, and \texttt{privacy\_id\_extractor}. 
For OpenDP, our participants found its API, including the metadata portion, difficult to use.

We recommend that DP tools should\textbf{carefully design APIs to obtain this metadata}, as well as assist users in configuring key DP-related metadata, including exposing metadata settings, providing documentation for each metadata setting, and auto-filling appropriate default values.

\subsection{Help Users with DP Foundations}
Our study surfaced a general need for additional resources to help data practitioners better understand DP concepts.
We found that many novices had difficulty understanding and describing the privacy budget (Section~\ref{sec:part-expl-dp}), and that both novices and experts sometimes had trouble describing the strength of the privacy guarantee (Section~\ref{results:RQ3:qual}). These results reinforced previous findings that DP concepts are complex and difficult to communicate~\cite{bullek2017towards, cummings2021need, xiong2020towards, kuhtreiber2022replication}, which inspire the following recommendations to address the challenge.

\paragraph{Provide general educational materials.}
Section~\ref{sec:RQ1:useful_source} suggests that our study instrument boosted participants' DP understanding.
DP tools may be able to replicate this effect by providing or directing users to general DP educational materials, similar to the handout and tutorials in this study.

\paragraph{Support privacy guarantee communication.}
Our participants had difficulty explaining the strength of the privacy guarantees, and several participants were unsure if their DP outputs would be private enough to be shared or published.
We encourage DP tools to provide users additional community resources~\cite{akil2017usability} on privacy guarantee (e.g., how to communicate the guarantee when disseminating DP analyses.)

\section{Conclusion}
We presented the first comprehensive usability study that evaluates four open-source Python-based DP tools with data practitioners.
Our findings suggest that DP tools should provide easy-to-navigate, DP-specific documentation, enhance error prevention and recovery capabilities, improve API designs to ease users' learning curves, and offer resources to strengthen users' DP foundations.
We aim for our findings and recommendations to facilitate broader DP adoption.

\section*{Acknowledgments}

The authors thank the SOUPS reviewers and shepherd for their hepful suggestions that resulted in significant improvements to the paper. This work was supported in part by an Amazon Research Award.

\bibliographystyle{plain}
\bibliography{usenix2024_SOUPS,dp}

\begin{thebibliography}{10}

\bibitem{acar2016you}
Yasemin Acar, Sascha Fahl, and Michelle~L Mazurek.
\newblock You are not your developer, either: A research agenda for usable security and privacy research beyond end users.
\newblock {\em 2016 IEEE Cybersecurity Development (SecDev)}, pages 3--8, 2016.

\bibitem{aghajani2020software}
Emad Aghajani, Csaba Nagy, Mario Linares-V{\'a}squez, Laura Moreno, Gabriele Bavota, Michele Lanza, and David~C Shepherd.
\newblock Software documentation: the practitioners' perspective.
\newblock In {\em Proceedings of the ACM/IEEE 42nd International Conference on Software Engineering}, pages 590--601, 2020.

\bibitem{aghajani2019software}
Emad Aghajani, Csaba Nagy, Olga~Lucero Vega-M{\'a}rquez, Mario Linares-V{\'a}squez, Laura Moreno, Gabriele Bavota, and Michele Lanza.
\newblock Software documentation issues unveiled.
\newblock In {\em 2019 IEEE/ACM 41st International Conference on Software Engineering (ICSE)}, pages 1199--1210. IEEE, 2019.

\bibitem{akil2017usability}
Bilal Akil, Ying Zhou, and Uwe R{\"o}hm.
\newblock On the usability of hadoop mapreduce, apache spark \& apache flink for data science.
\newblock In {\em 2017 IEEE International Conference on Big Data (Big Data)}, pages 303--310. IEEE, 2017.

\bibitem{usabilitytest07}
Morten~Sieker Andreasen, Henrik~Villemann Nielsen, Simon~Ormholt Schr\o{}der, and Jan Stage.
\newblock What happened to remote usability testing? an empirical study of three methods.
\newblock CHI '07, page 1405–1414, New York, NY, USA, 2007. Association for Computing Machinery.

\bibitem{appledpuse}
Apple.
\newblock {Apple: Differential Privacy Overview}, 2023.
\newblock \url{https://www.apple.com/privacy/docs/Differential_Privacy_Overview.pdf}.

\bibitem{ashenaSP2024}
Narges Ashena, Oana Inel, Badrie~L. Persaud, and Abraham Bernstein.
\newblock Casual users and rational choices within differential privacy.
\newblock In {\em Proceedings of the 2024 IEEE Symposium on Security and Privacy}, pages 88--88, 2024.

\bibitem{barik2017}
Titus Barik, Justin Smith, Kevin Lubick, Elisabeth Holmes, Jing Feng, Emerson Murphy-Hill, and Chris Parnin.
\newblock Do developers read compiler error messages?
\newblock In {\em 2017 IEEE/ACM 39th International Conference on Software Engineering (ICSE)}, pages 575--585, 2017.

\bibitem{becker2016}
Brett~A. Becker.
\newblock An effective approach to enhancing compiler error messages.
\newblock In {\em Proceedings of the 47th ACM Technical Symposium on Computing Science Education}, SIGCSE '16, page 126–131, New York, NY, USA, 2016. Association for Computing Machinery.

\bibitem{becker2019compiler}
Brett~A Becker, Paul Denny, Raymond Pettit, Durell Bouchard, Dennis~J Bouvier, Brian Harrington, Amir Kamil, Amey Karkare, Chris McDonald, Peter-Michael Osera, et~al.
\newblock Compiler error messages considered unhelpful: The landscape of text-based programming error message research.
\newblock {\em Proceedings of the working group reports on innovation and technology in computer science education}, pages 177--210, 2019.

\bibitem{berghel2022tumult}
Skye Berghel, Philip Bohannon, Damien Desfontaines, Charles Estes, Sam Haney, Luke Hartman, Michael Hay, Ashwin Machanavajjhala, Tom Magerlein, Gerome Miklau, et~al.
\newblock Tumult analytics: a robust, easy-to-use, scalable, and expressive framework for differential privacy.
\newblock {\em arXiv preprint arXiv:2212.04133}, 2022.

\bibitem{bevan2006practical}
Nigel Bevan.
\newblock Practical issues in usability measurement.
\newblock {\em Interactions}, 13(6):42--43, 2006.

\bibitem{brooke1996sus}
John Brooke.
\newblock Sus: a “quick and dirty’usability.
\newblock {\em Usability evaluation in industry}, 189(3), 1996.

\bibitem{bullek2017towards}
Brooke Bullek, Stephanie Garboski, Darakhshan~J Mir, and Evan~M Peck.
\newblock Towards understanding differential privacy: When do people trust randomized response technique?
\newblock In {\em Proceedings of the 2017 CHI Conference on Human Factors in Computing Systems}, pages 3833--3837, 2017.

\bibitem{carlini2019secret}
Nicholas Carlini, Chang Liu, {\'U}lfar Erlingsson, Jernej Kos, and Dawn Song.
\newblock The secret sharer: Evaluating and testing unintended memorization in neural networks.
\newblock In {\em 28th {USENIX} Security Symposium ({USENIX} Security 19)}, pages 267--284, 2019.

\bibitem{carlini2021extracting}
Nicholas Carlini, Florian Tramer, Eric Wallace, Matthew Jagielski, Ariel Herbert-Voss, Katherine Lee, Adam Roberts, Tom Brown, Dawn Song, Ulfar Erlingsson, et~al.
\newblock Extracting training data from large language models.
\newblock In {\em 30th USENIX Security Symposium (USENIX Security 21)}, pages 2633--2650, 2021.

\bibitem{casacuberta2022widespread}
S{\'\i}lvia Casacuberta, Michael Shoemate, Salil Vadhan, and Connor Wagaman.
\newblock Widespread underestimation of sensitivity in differentially private libraries and how to fix it.
\newblock In {\em Proceedings of the 2022 ACM SIGSAC Conference on Computer and Communications Security}, pages 471--484, 2022.

\bibitem{cooke2010assessing}
Lynne Cooke.
\newblock Assessing concurrent think-aloud protocol as a usability test method: A technical communication approach.
\newblock {\em IEEE Transactions on Professional Communication}, 53(3):202--215, 2010.

\bibitem{cummings2021need}
Rachel Cummings, Gabriel Kaptchuk, and Elissa~M Redmiles.
\newblock " i need a better description": An investigation into user expectations for differential privacy.
\newblock In {\em Proceedings of the 2021 ACM SIGSAC Conference on Computer and Communications Security}, pages 3037--3052, 2021.

\bibitem{desfontaines2020lowering}
Damien Desfontaines.
\newblock {\em Lowering the cost of anonymization}.
\newblock PhD thesis, ETH Zurich, 2020.

\bibitem{diffprivlib}
{DiffPrivLib}, 2023.
\newblock \url{https://github.com/IBM/differential-privacy-library}.

\bibitem{dpcreator}
{DP Creator}, 2023.
\newblock \url{https://github.com/opendp/dpcreator}.

\bibitem{dumas1999practical}
Joseph~S Dumas and Janice Redish.
\newblock A practical guide to usability testing, 1999.

\bibitem{dwork2006calibrating}
Cynthia Dwork, Frank McSherry, Kobbi Nissim, and Adam Smith.
\newblock Calibrating noise to sensitivity in private data analysis.
\newblock In {\em Theory of cryptography conference}, pages 265--284. Springer, 2006.

\bibitem{dwork2014algorithmic}
Cynthia Dwork, Aaron Roth, et~al.
\newblock The algorithmic foundations of differential privacy.
\newblock {\em Foundations and Trends{\textregistered} in Theoretical Computer Science}, 9(3--4):211--407, 2014.

\bibitem{fereday2006demonstrating}
Jennifer Fereday and Eimear Muir-Cochrane.
\newblock Demonstrating rigor using thematic analysis: A hybrid approach of inductive and deductive coding and theme development.
\newblock {\em International journal of qualitative methods}, 5(1):80--92, 2006.

\bibitem{gaboardi2016psi}
Marco Gaboardi, James Honaker, Gary King, Jack Murtagh, Kobbi Nissim, Jonathan Ullman, and Salil Vadhan.
\newblock Psi ($\{$$\backslash$Psi$\}$): a private data sharing interface.
\newblock {\em arXiv preprint arXiv:1609.04340}, 2016.

\bibitem{garrido2022lessons}
Gonzalo~M Garrido, Xiaoyuan Liu, Florian Matthes, and Dawn Song.
\newblock Lessons learned: Surveying the practicality of differential privacy in the industry.
\newblock {\em Proceedings on Privacy Enhancing Technologies}, 2:151--170, 2023.

\bibitem{googledpuse}
Google.
\newblock {Google: Differentially private heatmaps}, 2023.
\newblock \url{https://blog.research.google/2023/04/differentially-private-heatmaps.html}.

\bibitem{googledp}
{Google's differential privacy libraries}, 2023.
\newblock \url{https://github.com/google/differential-privacy}.

\bibitem{grisaffe2007questions}
Douglas~B Grisaffe.
\newblock Questions about the ultimate question: conceptual considerations in evaluating reichheld's net promoter score (nps).
\newblock {\em Journal of Consumer Satisfaction, Dissatisfaction and Complaining Behavior}, 20:36, 2007.

\bibitem{haney2022precision}
Samuel Haney, Damien Desfontaines, Luke Hartman, Ruchit Shrestha, and Michael Hay.
\newblock Precision-based attacks and interval refining: how to break, then fix, differential privacy on finite computers.
\newblock {\em arXiv preprint arXiv:2207.13793}, 2022.

\bibitem{Hartmann2010}
Bj\"{o}rn Hartmann, Daniel MacDougall, Joel Brandt, and Scott~R. Klemmer.
\newblock What would other programmers do: suggesting solutions to error messages.
\newblock CHI '10, page 1019–1028, New York, NY, USA, 2010. Association for Computing Machinery.

\bibitem{jayaraman2020revisiting}
Bargav Jayaraman, Lingxiao Wang, Katherine Knipmeyer, Quanquan Gu, and David Evans.
\newblock Revisiting membership inference under realistic assumptions.
\newblock {\em arXiv preprint arXiv:2005.10881}, 2020.

\bibitem{jin2022we}
Jiankai Jin, Eleanor McMurtry, Benjamin~IP Rubinstein, and Olga Ohrimenko.
\newblock Are we there yet? timing and floating-point attacks on differential privacy systems.
\newblock In {\em 2022 IEEE Symposium on Security and Privacy (SP)}, pages 473--488. IEEE, 2022.

\bibitem{johnson2020chorus}
Noah Johnson, Joseph~P Near, Joseph~M Hellerstein, and Dawn Song.
\newblock Chorus: a programming framework for building scalable differential privacy mechanisms.
\newblock In {\em 2020 IEEE European Symposium on Security and Privacy (EuroS\&P)}, pages 535--551. IEEE, 2020.

\bibitem{kifer2020guidelines}
Daniel Kifer, Solomon Messing, Aaron Roth, Abhradeep Thakurta, and Danfeng Zhang.
\newblock Guidelines for implementing and auditing differentially private systems.
\newblock {\em arXiv preprint arXiv:2002.04049}, 2020.

\bibitem{kuhtreiber2022replication}
Patrick K{\"u}htreiber, Viktoriya Pak, and Delphine Reinhardt.
\newblock Replication: The effect of differential privacy communication on german users' comprehension and data sharing attitudes.
\newblock In {\em Eighteenth Symposium on Usable Privacy and Security (SOUPS 2022)}, pages 117--134, 2022.

\bibitem{li2018coconut}
Tianshi Li, Yuvraj Agarwal, and Jason~I Hong.
\newblock Coconut: An ide plugin for developing privacy-friendly apps.
\newblock {\em Proceedings of the ACM on Interactive, Mobile, Wearable and Ubiquitous Technologies}, 2(4):1--35, 2018.

\bibitem{databreach2}
Wired Magazine.
\newblock {T-Mobile's \$150 Million Security Plan Isn't Cutting It}, 2023.
\newblock \url{https://www.wired.com/story/tmobile-data-breach-again/}.

\bibitem{mcsherry2009privacy}
Frank~D McSherry.
\newblock Privacy integrated queries: an extensible platform for privacy-preserving data analysis.
\newblock In {\em Proceedings of the 2009 ACM SIGMOD International Conference on Management of data}, pages 19--30, 2009.

\bibitem{mehta2016comparative}
Parmita Mehta, Sven Dorkenwald, Dongfang Zhao, Tomer Kaftan, Alvin Cheung, Magdalena Balazinska, Ariel Rokem, Andrew Connolly, Jacob Vanderplas, and Yusra AlSayyad.
\newblock Comparative evaluation of big-data systems on scientific image analytics workloads.
\newblock {\em arXiv preprint arXiv:1612.02485}, 2016.

\bibitem{meng2018application}
Michael Meng, Stephanie Steinhardt, and Andreas Schubert.
\newblock Application programming interface documentation: What do software developers want?
\newblock {\em Journal of Technical Writing and Communication}, 48(3):295--330, 2018.

\bibitem{microsoftdpuse}
Microsoft.
\newblock {Microsoft AI: Differential Privacy}, 2023.
\newblock \url{https://www.microsoft.com/en-us/ai/ai-lab-differential-privacy}.

\bibitem{mironov2012significance}
Ilya Mironov.
\newblock On significance of the least significant bits for differential privacy.
\newblock In {\em Proceedings of the 2012 ACM conference on Computer and communications security}, pages 650--661, 2012.

\bibitem{murtagh2018usable}
Jack Murtagh, Kathryn Taylor, George Kellaris, and Salil Vadhan.
\newblock Usable differential privacy: A case study with psi.
\newblock {\em arXiv preprint arXiv:1809.04103}, 2018.

\bibitem{myllarniemi2018development}
Varvana Myll{\"a}rniemi, Sari Kujala, Mikko Raatikainen, and Piia Sevo{\'n}n.
\newblock Development as a journey: factors supporting the adoption and use of software frameworks.
\newblock {\em Journal of software engineering research and development}, 6:1--22, 2018.

\bibitem{naiakshina2017developers}
Alena Naiakshina, Anastasia Danilova, Christian Tiefenau, Marco Herzog, Sergej Dechand, and Matthew Smith.
\newblock Why do developers get password storage wrong? a qualitative usability study.
\newblock In {\em Proceedings of the 2017 ACM SIGSAC Conference on Computer and Communications Security}, pages 311--328, 2017.

\bibitem{nanayakkara2022visualizing}
Priyanka Nanayakkara, Johes Bater, Xi~He, Jessica Hullman, and Jennie Rogers.
\newblock Visualizing privacy-utility trade-offs in differentially private data releases.
\newblock {\em Proceedings on Privacy Enhancing Technologies}, 2022(2):601--618.

\bibitem{nielsen1994usability}
Jakob Nielsen.
\newblock {\em Usability engineering}.
\newblock Morgan Kaufmann, 1994.

\bibitem{nielsen1996usability}
Jakob Nielsen.
\newblock Usability metrics: Tracking interface improvements.
\newblock {\em IEEE software}, 13(6):1--2, 1996.

\bibitem{opendp}
OpenDP, 2023.
\newblock \url{https://github.com/opendp/opendp}.

\bibitem{papernot2019machine}
Nicolas Papernot.
\newblock Machine learning at scale with differential privacy in {TensorFlow}.
\newblock In {\em 2019 {USENIX} Conference on Privacy Engineering Practice and Respect ({PEPR} 19)}, 2019.

\bibitem{pipelinedp}
{PipelineDP}, 2023.
\newblock \url{https://pipelinedp.io/}.

\bibitem{prather2017}
James Prather, Raymond Pettit, Kayla~Holcomb McMurry, Alani Peters, John Homer, Nevan Simone, and Maxine Cohen.
\newblock On novices' interaction with compiler error messages: A human factors approach.
\newblock In {\em Proceedings of the 2017 ACM Conference on International Computing Education Research}, ICER '17, page 74–82, New York, NY, USA, 2017. Association for Computing Machinery.

\bibitem{databreach1}
Associated Press.
\newblock {Wawa agrees to payment, security changes for '19 data breach}, 2022.
\newblock \url{https://apnews.com/article/technology-pennsylvania-malware-attorney-generals-office-0ebedd8dce8bf0e21833f52944a48b56}.

\bibitem{privacybeam}
{Privacy on Beam}, 2023.
\newblock \url{https://github.com/google/differential-privacy/tree/main/privacy-on-beam}.

\bibitem{chorusweb}
Chorus Repository, 2023.
\newblock \url{https://github.com/uvm-plaid/chorus}.

\bibitem{sarathy2023don}
Jayshree Sarathy, Sophia Song, Audrey Haque, Tania Schlatter, and Salil Vadhan.
\newblock Don’t look at the data! how differential privacy reconfigures the practices of data science.
\newblock In {\em Proceedings of the 2023 CHI Conference on Human Factors in Computing Systems}, pages 1--19, 2023.

\bibitem{shokri2017membership}
Reza Shokri, Marco Stronati, Congzheng Song, and Vitaly Shmatikov.
\newblock Membership inference attacks against machine learning models.
\newblock In {\em 2017 IEEE Symposium on Security and Privacy (SP)}, pages 3--18. IEEE, 2017.

\bibitem{govtechbenchmarks}
Anshu Singh and Syahri Ikram.
\newblock Benchmarking differential privacy python tools.
\newblock \url{https://github.com/dsaidgovsg/benchmarking-differential-privacy-tools}, 2023.

\bibitem{traver2010compiler}
V~Javier Traver.
\newblock On compiler error messages: what they say and what they mean.
\newblock {\em Advances in Human-Computer Interaction}, 2010:1--26, 2010.

\bibitem{treude2014extracting}
Christoph Treude, Martin~P Robillard, and Barth{\'e}l{\'e}my Dagenais.
\newblock Extracting development tasks to navigate software documentation.
\newblock {\em IEEE Transactions on Software Engineering}, 41(6):565--581, 2014.

\bibitem{censusdpuse}
{U.S. Census Bureau}.
\newblock {Why the Census Bureau Chose Differential Privacy}, 2023.
\newblock \url{https://www.census.gov/library/publications/2023/decennial/c2020br-03.html}.

\bibitem{wilson2020differentially}
Royce~J Wilson, Celia~Yuxin Zhang, William Lam, Damien Desfontaines, Daniel Simmons-Marengo, and Bryant Gipson.
\newblock Differentially private sql with bounded user contribution.
\newblock {\em Proceedings on Privacy Enhancing Technologies}, 2:230--250, 2020.

\bibitem{xiong2020towards}
Aiping Xiong, Tianhao Wang, Ninghui Li, and Somesh Jha.
\newblock Towards effective differential privacy communication for users’ data sharing decision and comprehension.
\newblock In {\em 2020 IEEE Symposium on Security and Privacy (SP)}, pages 392--410. IEEE, 2020.

\bibitem{xiong2022using}
Aiping Xiong, Chuhao Wu, Tianhao Wang, Robert~W Proctor, Jeremiah Blocki, Ninghui Li, and Somesh Jha.
\newblock Using illustrations to communicate differential privacy trust models: An investigation of users' comprehension, perception, and data sharing decision.
\newblock {\em arXiv preprint arXiv:2202.10014}, 2022.

\bibitem{yousefpour2021opacus}
Ashkan Yousefpour, Igor Shilov, Alexandre Sablayrolles, Davide Testuggine, Karthik Prasad, Mani Malek, John Nguyen, Sayan Ghosh, Akash Bharadwaj, Jessica Zhao, et~al.
\newblock Opacus: User-friendly differential privacy library in pytorch.
\newblock {\em arXiv preprint arXiv:2109.12298}, 2021.

\bibitem{zetasql}
{ZetaSQL differential privacy extension}, 2023.
\newblock \url{https://github.com/google/differential-privacy/tree/main/examples/zetasql}.

\end{thebibliography}

\appendix
\section{Eligibility Survey}
\label{sec:eligibility-survey}
\textbf{For questions that test participants' understanding, we highlight the correct answer in bold.}

\paragraph{Eligibility Questions after displaying IRB-approved consent form}

\begin{itemize}
    \item I have read and understood the information above. No/Yes
    \item I want to proceed to complete the eligibility survey for this research study. No/Yes
    \item Are you at least 18 years old? No/Yes
    \item Do you reside in the United States? No/Yes
    \item Have you performed statistical data analysis in Python? No/Yes
    \item Have you used the Jupyter Notebook before? No/Yes
    \item Are you willing to participate in a study to evaluate a data science tool that will require you to code in Python in a Jupyter Notebook? No/Yes
    \item Are you willing to participate in a 1.5-hour usability study remotely via Microsoft Teams? No/Yes
\end{itemize}

\paragraph{Questions on Python, DP, and basic demographics}

\begin{enumerate}
    \item How  many years have you been coding in Python?
    \begin{enumerate}
        \item 0-1
        \item 2-3
        \item More than 3
    \end{enumerate}
    \item How many years have you been using the Jupyter Notebook?
    \begin{enumerate}
        \item 0-1
        \item 2-3
        \item More than 3
    \end{enumerate}
    \item Which of the following best describes how you use Python and the Jupyter notebook for statistical analysis?
    \begin{enumerate}
        \item They are my preferred language/tool
        \item I am comfortable using them but I prefer other languages/tools (e.g., R)
        \item I can work with them but often need to resort to documentation
        \item I rarely use them and need additional time to get familiar with them.
    \end{enumerate}
    \item Use “set” instead of “list” as a Python data structure for a sequence of elements when:
    \begin{enumerate}
        \item elements will be appended to increase the size of the sequence
        \item  the order of items is important
        \item \textbf{it is important to know if the sequence contains a specific item}
        \item it is important to know the item with maximum value in the sequence
        \item I don't know
    \end{enumerate}
   \item 
        What is the output of the following code? 
\begin{verbatim}
str1 = "DataScience is fun!" 
print(str1[4:12])
\end{verbatim}
   \begin{enumerate}
       \item  \textbf{Science}
        \item Data Sci
        \item aScience
        \item Error
        \item I don't know
   \end{enumerate}

\item Have you heard of the term differential privacy (DP) before?
\begin{enumerate}
\item No 
\item Yes
\end{enumerate}

\item Have you ever written code to implement differential privacy (DP) in any capacity?
\begin{enumerate}
\item No 
\item Yes
\end{enumerate}

\item In differential privacy, which value of the privacy parameter $\epsilon$ provides stronger privacy?
\begin{enumerate}
\item \textbf{$\epsilon$ = 0.1}
\item $\epsilon$ = 1.0
\item I don't know
\end{enumerate}

\item Releasing two differentially private statistics, one with $\epsilon_1$ = 0.1 and the other with $\epsilon_2$ = 0.5, results in a total privacy loss of:
\begin{enumerate}
\item $\epsilon$ = 0.1
\item $\epsilon$ = 0.5
\item \textbf{$\epsilon$ = 0.6}
\item $\epsilon$ = 0.05
\item I don't know
\end{enumerate}

\item If the mechanism M returns a number and satisfies differential privacy with $\epsilon$ = 0.1, does abs(M(x)) satisfy differential privacy, where abs is the absolute value function?
\begin{enumerate}
\item No, not necessarily
\item \textbf{Yes, for $\epsilon$ = 0.1}
\item Yes, for some $\epsilon$ > 0.1
\item I don't know
\end{enumerate}

\item Which of the following is an advantage of using Differential Privacy?
\begin{enumerate}
\item It guarantees complete anonymity of the data subjects
\item It ensures that the data is completely accurate
\item \textbf{It provides a tradeoff between privacy and utility of the data}
\item It is a computationally simple method for preserving privacy in large datasets
\item I don't know
\end{enumerate}

\item What is your age?

\item What is your gender?

\item Are you an undergraduate or a graduate student?
\end{enumerate}

\section{The Handout and the Tutorials}
\label{sec:handout}
Available at Open Science Framework (OSF):
\url{ https://osf.io/ag2fj/?view_only=29a9bc2a30574befa9f3d0643951b9c6} 

\section{Post-Task Survey}
\label{sec:post-task-survey}

\textbf{For questions that test participants' understanding, we highlight the correct answer in bold.}
\begin{enumerate}
\item Please enter your participant ID

\item Please rate the following statements using the [Likert] scale indicated below.
  \begin{enumerate}
  \item I think that I would like to use [DP tool] frequently.
  \item I found [DP tool] unnecessarily complex.
  \item I thought [DP tool] was easy to use.
  \item I think that I would need the support of a technical person to be able to use [DP tool].
  \item I found the various functions in [DP tool] were well integrated.
  \item I thought there was too much inconsistency in [DP tool].
  \item I would imagine that most people would learn to use [DP tool] very quickly.
  \item I found [DP tool] very cumbersome to use.
  \item I felt very confident using [DP tool].
  \item I needed to learn a lot of things before I could get going with [DP tool].
  \item I found [DP tool] introduced DP concepts appropriately for me to perform the tasks.
  \item I feel I have to learn DP concepts more systematically to solve the tasks.
  \end{enumerate}

\item If another data scientist that you know needs to use differential privacy in their data analysis, how likely is it that you would recommend Tumult Analytics to them?
  \begin{itemize}
  \item 10-point Likert scale, "Not at all likely" to "Extremely likely"
  \end{itemize}

\item If you completed at least one task in the study, what helped you successfully complete the task(s)? Choose all that apply.
  \begin{enumerate}
\item The Differential Privacy handout (including the video)
\item The [DP tool] tutorial (including its examples)
\item The official [DP tool] documentation
\item My prior data science skills (like Python, Pandas, statistics, etc.)
\item My prior knowledge of Differential Privacy
\item Other (please specify)
\item N/A (I didn't complete any tasks)
  \end{enumerate}

\item What hindered your completion of the tasks? Choose all that apply.
  \begin{enumerate}
  \item The Differential Privacy handout (including the video)
  \item The [DP tool] tutorial (including its examples)
  \item The official [DP tool] documentation
  \item My prior data science skills (like Python, Pandas, statistics, etc.)
  \item My prior knowledge of Differential Privacy
  \item Other (please specify)
  \end{enumerate}

\item If the mechanism M returns a number and satisfies differential privacy with $\epsilon$ = 0.1, does abs(M(x)) satisfy differential privacy, where abs is the absolute value function?
  \begin{enumerate}
  \item No, not necessarily
  \item \textbf{Yes, for $\epsilon$ = 0.1}
  \item Yes, for some $\epsilon$ > 0.1
  \item I don't know
  \end{enumerate}

\item In differential privacy, which value of the privacy parameter $\epsilon$ provides stronger privacy?
  \begin{enumerate}
  \item \textbf{$\epsilon$ = 0.1}
  \item $\epsilon$ = 1.0
  \item I don't know
  \end{enumerate}

\item Releasing two differentially private statistics, one with $\epsilon_1$ = 0.1 and the other with $\epsilon_2$ = 0.5, results in a total privacy loss of:
  \begin{enumerate}
  \item $\epsilon$ = 0.1
  \item $\epsilon$ = 0.5
  \item \textbf{$\epsilon$ = 0.6}
  \item $\epsilon$ = 0.05
  \item I don't know
  \end{enumerate}

\item Which of the following is an advantage of using Differential Privacy?
  \begin{enumerate}
  \item It guarantees complete anonymity of the data subjects
  \item It ensures that the data is completely accurate
  \item \textbf{It provides a tradeoff between privacy and utility of the data}
  \item It is a computationally simple method for preserving privacy in large datasets
  \item I don't know
  \end{enumerate}
\end{enumerate}

\section{Post-Task Interview}
\label{sec:post-task-interview}

\textbf{For questions that test participants' understanding, we give sample correct answers after each question.}

Thank you for completing/making an effort to complete the tasks with [tool] and the post-task
survey. Now we have a few questions for you to reflect on your experience with the study.
\begin{enumerate}
  \item After completing the tasks, can you explain differential privacy to me in your own words?\\
  \textbf{Correct answer:} Differential privacy is a formal property that limits the distributional difference between a statistic computed on one dataset and the same statistic computed on a neighboring dataset.
  \item Consider the tasks you worked on just now, can you explain:
  \begin{enumerate}
      \item What was the privacy budget for each task? \\
      \textbf{Correct answer:} Depends on the parameters used by the participant---it should be equal to the value of $\epsilon$ used in each task's solution.
      \item What was Epsilon?\\
      \textbf{Correct answer:} Same as (a).
      \item What was the total privacy budget for the whole notebook?
      \\
      \textbf{Correct answer:} 3 $\times$ (answer from (a)), by sequential composition.
      \item If the results you computed were released to the public, how strong would you expect the privacy protection for individuals in the original data to be?
  \end{enumerate}
  
  \item During this study, what helped you most in understanding the concepts (e.g., privacy budget, Epsilon) that we discussed just now? Please rank the following options from “most useful” to “least useful”. 
  \begin{enumerate}
      \item The Differential Privacy handout (including the video)
      \item The [DP tool] tutorial (including its examples)
      \item The official [DP tool] documentation
      \item My prior knowledge of Differential Privacy
      \item Other (please specify)
  \end{enumerate}

  \item When using [tool] in the study, what aspects/components of [tool] do you think are helpful for you to complete the tasks?
  \item When using [tool] in the study, what aspects/components of [tool] do you think are frustrating for you to complete the tasks?
  \item After this study, what recommendation(s) do you have to improve the usability of [tool]?
  \item Can you tell us what helped you successfully complete the task(s)? 
  \begin{enumerate}
        \item The Differential Privacy handout (including the video)
        \item The [DP tool] tutorial (including its examples)
        \item The official [DP tool] documentation
        \item My prior data science skills (like Python, Pandas, statistics, etc.)
        \item My prior knowledge of Differential Privacy
        \item Other (please specify)
  \end{enumerate}
  
  \item Can you tell us what hindered your completion of the (task)?
  \begin{enumerate}
      
        \item The Differential Privacy handout (including the video)
        \item The [DP tool] tutorial (including its examples)
        \item The official [DP tool] documentation
        \item My prior data science skills (like Python, Pandas, statistics, etc.)
        \item My prior knowledge of Differential Privacy
        \item Other (please specify)
  \end{enumerate}
\end{enumerate}

\section{Task Dataset}
\label{app:data}

The dataset used for the tasks was provided to participants in a CSV file. This is a synthetic dataset that counted restaurant visits across a week, where each record represented a distinct visit with a visitor ID. 
The full dataset is available at Open Science Framework (OSF):
\url{ https://osf.io/ag2fj/?view_only=29a9bc2a30574befa9f3d0643951b9c6} 



\section{Task Solutions}
\label{app:task-solutions}

We wrote sample solutions for the three tasks from Table~\ref{tab:usability_tasks} for each tool we studied. Participant solutions were usually similar, but not necessarily identical. We will make these sample solutions available publicly via Open Science Framework on publication of the paper.

\section{Additional Figures}
\label{sec:additional-results}

\begin{figure*}
    \centering
    \begin{subfigure}[b]{0.6\textwidth}
        \centering 
        \includegraphics[width=\textwidth]{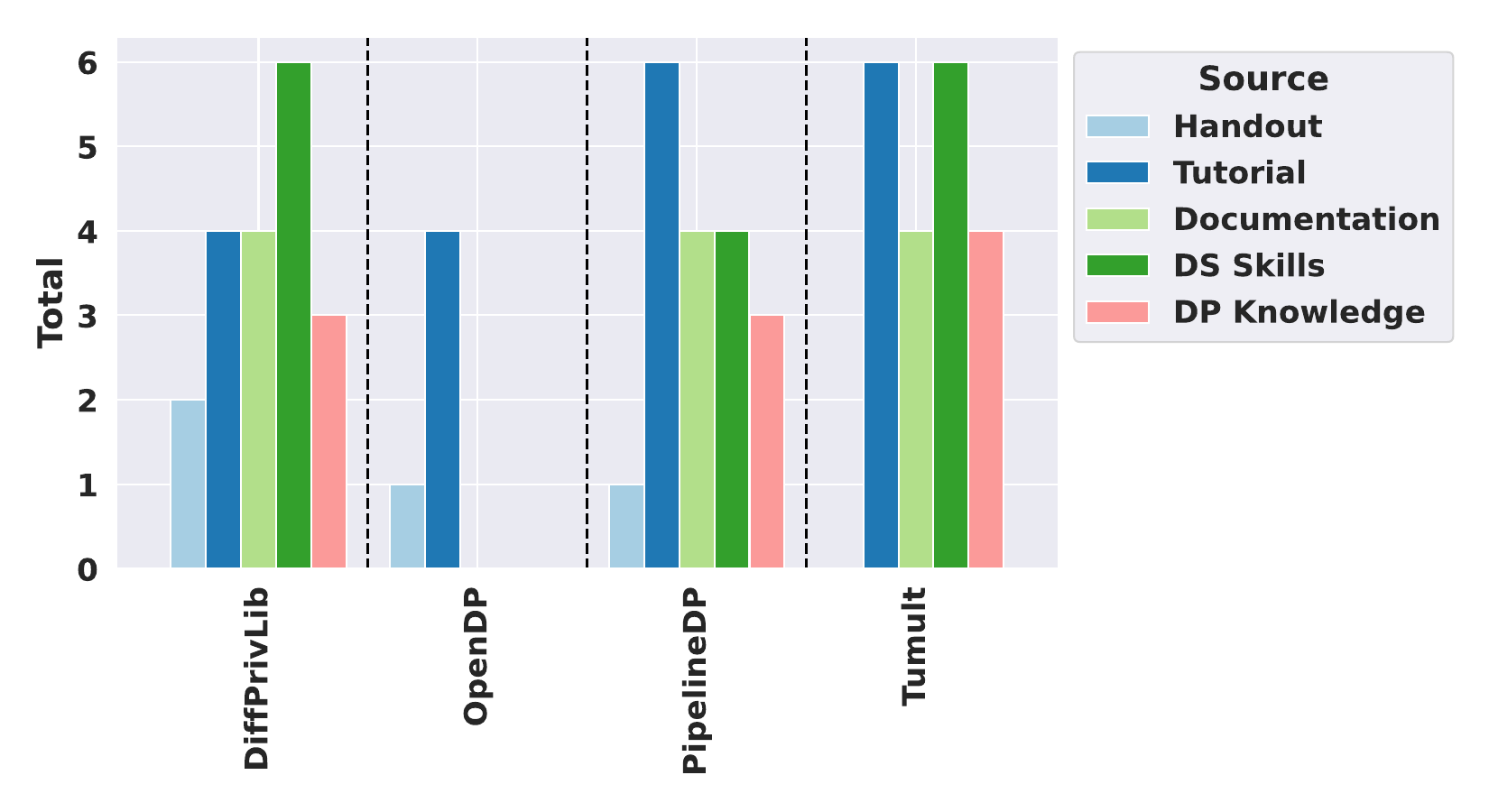}
        \caption{By Tool}
        \label{subfig:factors_helping_tool}
    \end{subfigure}%
    \hfill 
    \begin{subfigure}[b]{0.4\textwidth} 
        \centering
        \includegraphics[width=\textwidth]{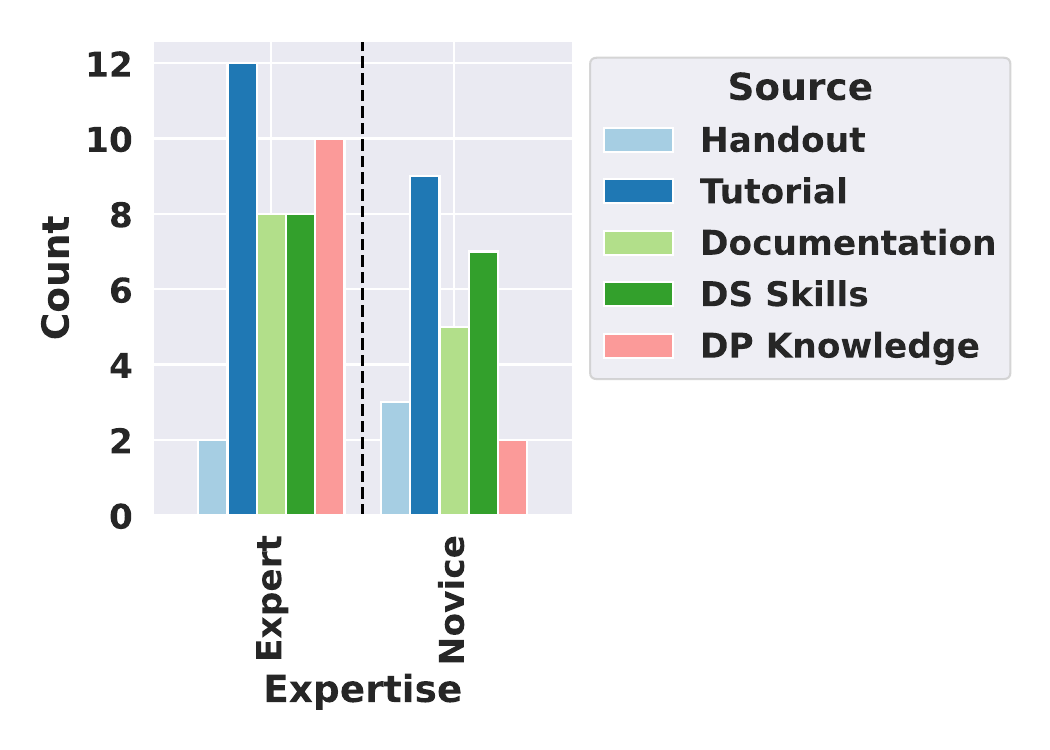}
        \caption{By Expertise}
        \label{subfig:factors_hindering_tool}
    \end{subfigure}%
    \caption{Factors helping task completion by tool and expertise.} 
    \label{fig:factors_helping}
\end{figure*}

\begin{figure*}
    \centering
    \begin{subfigure}[b]{0.6\textwidth}
        \centering 
        \includegraphics[width=\textwidth]{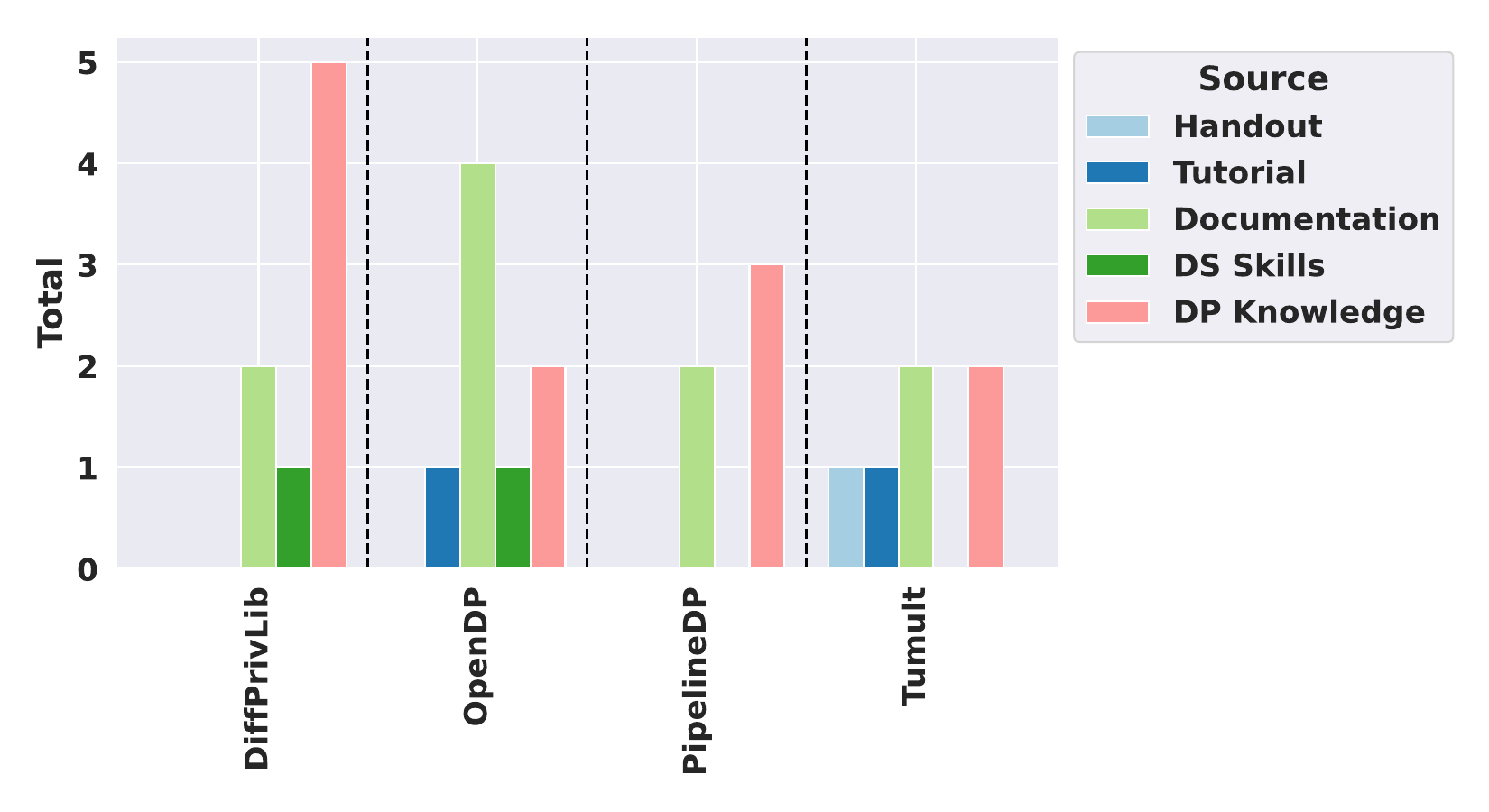}
        \caption{By Tool}
        \label{subfig:factors_helping_expertise}
    \end{subfigure}%
    \hfill 
    \begin{subfigure}[b]{0.4\textwidth} 
        \centering
        \includegraphics[width=\textwidth]{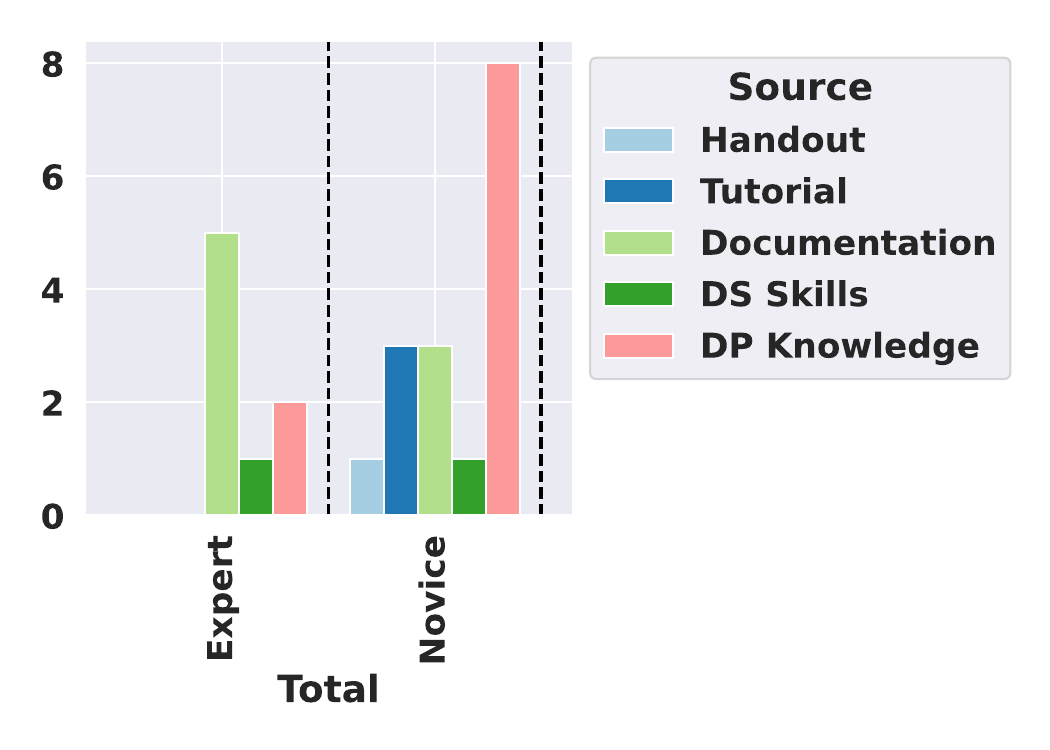}
        \caption{By Expertise}
        \label{subfig:factors_hindering_expertise}
    \end{subfigure}%
    \caption{Factors hindering task completion by tool and expertise.}
    \label{fig:factors_hindering}
\end{figure*}

\begin{figure*}
    \centering
    \begin{subfigure}[b]{0.55\textwidth} 
        \centering
        \includegraphics[width=\textwidth]{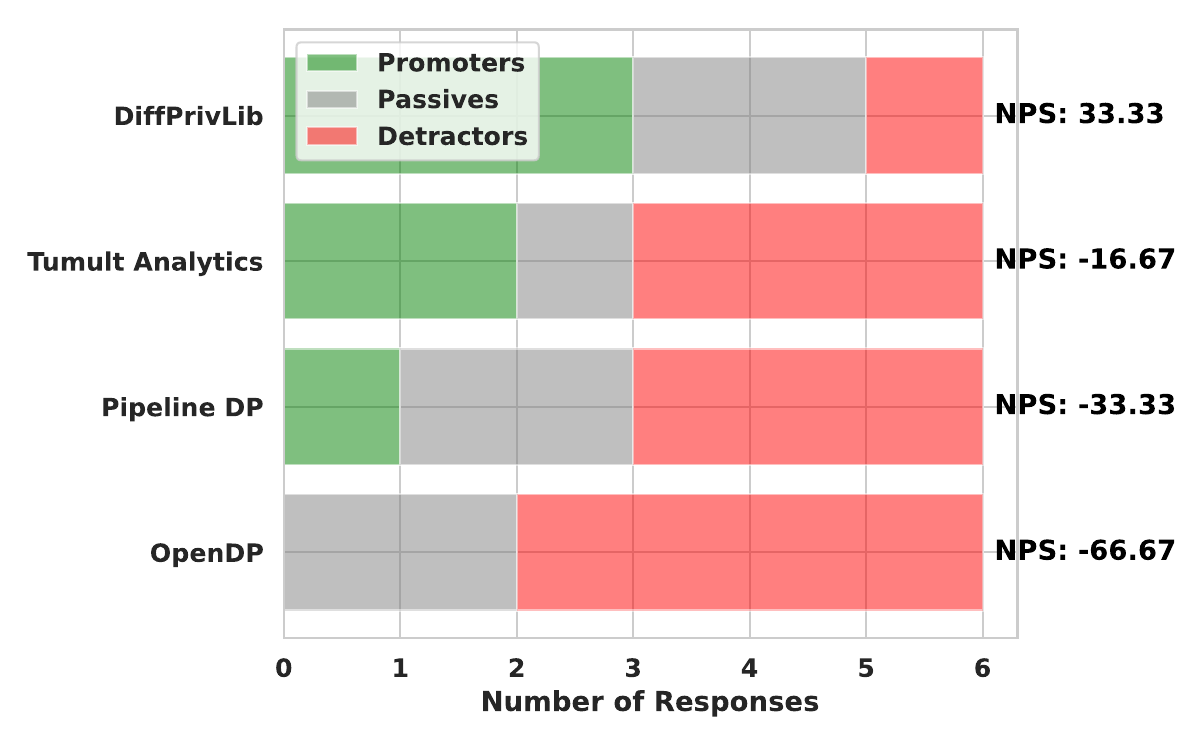}
        \caption{Net Promoter Score (NPS) results} 
        \label{subfig:nps}
    \end{subfigure}%
    \hfill 
    \begin{subfigure}[b]{0.45\textwidth}
        \centering 
        \includegraphics[width=\textwidth]{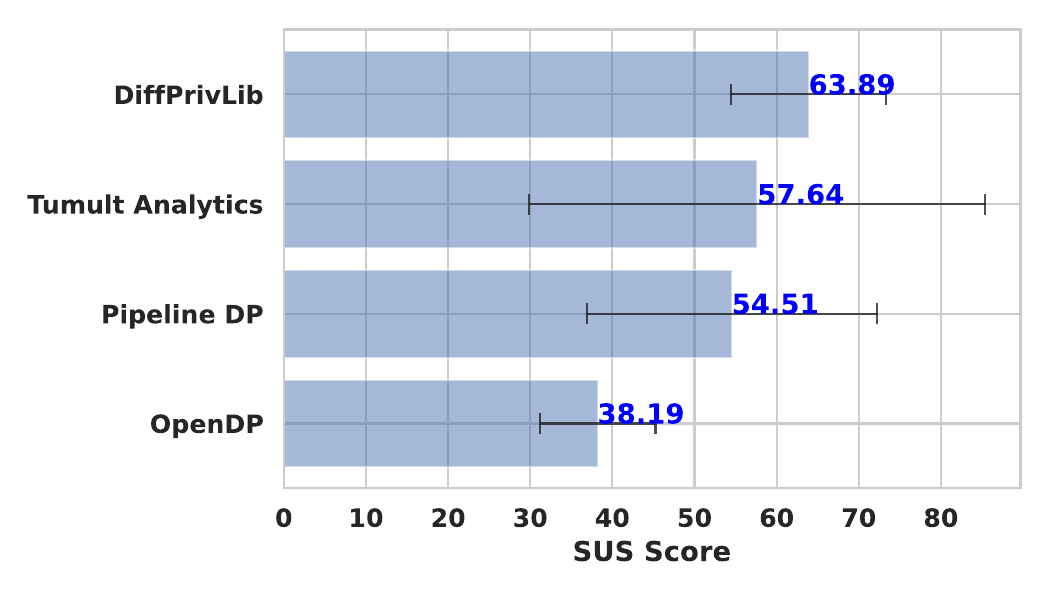}
        \caption{System Usability Scale (SUS) results}
        \label{subfig:sus}
    \end{subfigure}
    \caption{User satisfaction scores: (a) Net Promoter Score (NPS), and (b) System Usability Score (SUS).}
    \label{fig:satisfaction}
\end{figure*}

\end{document}